\documentclass[%
showkeys,
reprint,
superscriptaddress,
showpacs,
preprintnumbers,
nofootinbib,
amsmath,
amssymb,
aps,
prd,
tightenlines,
11pt,
notitlepage
]{revtex4-1}
\usepackage[T1]{fontenc}

\usepackage{alltt}
\usepackage{amsmath}
\usepackage{amsthm}
\usepackage{amssymb}
\usepackage{bm}
\usepackage{booktabs,siunitx}
\usepackage{cancel}
\usepackage{color}
\usepackage{csquotes}
\usepackage{enumerate}
\usepackage{graphicx,bm}
\usepackage{lipsum}
\usepackage{listings}
\usepackage{mathtools}
\usepackage{mathrsfs}
\usepackage{physics}
\usepackage{relsize}
\usepackage{slashed}
\usepackage{soul}
\usepackage{spverbatim}
\usepackage{wrapfig}
\usepackage{xcolor}
\usepackage{diagbox}
\usepackage{multirow}
\usepackage{subcaption}
\usepackage[utf8]{inputenc}
\usepackage[pdftex,hypertexnames=false,linktocpage=true]{hyperref}
\hypersetup{colorlinks=true,linkcolor=blue,anchorcolor=blue,citecolor=darkgreen,filecolor=blue,urlcolor=blue,bookmarksnumbered=true,
	pdfview=FitB
}
\colorlet{darkgreen}{green!60!black}
\colorlet{brightyellow}{yellow!75!red}
\colorlet{orange}{red!50!yellow}
\colorlet{darkblue}{blue!60!black}
\colorlet{darkred}{red!80!black}
\colorlet{greenblue}{green!50!blue}

\makeatletter

\newcommand{\Rmnum}[1]{\expandafter\@slowromancap\romannumeral #1@}
\makeatother

\def\dd{{\mathrm{d}}}
\def\imag{{\mathrm{i}}}

\begin{document}
	
	\title{Semileptonic Decay of $B_c$ to $\eta_c$ and $J/\psi$ on the Light Front}
	
	\author{Shuo Tang}
	\affiliation{%
		Department of Physics and Astronomy, Iowa State University, Ames, IA 50011, USA
	}%
	
	\author{Shaoyang Jia}
	\email[Corresponding author: ] {syjia@anl.gov}
	\affiliation{%
		Department of Physics and Astronomy, Iowa State University, Ames, IA 50011, USA
	}	
	\affiliation{%
		Physics Division, Argonne National Laboratory, Lemont, IL 60439, USA
	}	
	
	\author{Pieter Maris}
	\affiliation{%
		Department of Physics and Astronomy, Iowa State University, Ames, IA 50011, USA
	}
	
	\author{James P. Vary}
	\affiliation{%
		Department of Physics and Astronomy, Iowa State University, Ames, IA 50011, USA
	}
	\date{\today}
	
	\begin{abstract}
		We study the semileptonic decay of the $B_c (0^-)$ to charmonia through the bottom-to-charm-quark electroweak current in the framework of basis light-front quantization.
		Explicitly, we calculate the weak transition form factors for processes of $B_c$ decaying into $\eta_c$ or $J/\psi$ based on the corresponding initial and final valence light-front wave functions both obtained from the basis light-front quantization. 
		We also present the corresponding differential decay width and branching ratios, as well as the branching ratios for decays into the $\eta_c'$ and the $\psi'$.
		We observe unphysical frame dependence of the calculated form factors, which is attributed to only including the valence light-front wave functions. Based on the analysis of current component, we propose a preferred set of frames to calculate these form factors.
		
	\end{abstract}
	\maketitle
	
	\section{Introduction}

	Precise measurement of elements ${V_{q_1 q_2}}$ in the Cabibbo-Kobayashi-Maskawa (CKM) matrix is crucial to the electroweak theory. Among these elements, the determination of ${V_{ub}}$ and ${V_{cb}}$ leads a central test of the Standard Model for heavy-flavor physics. Although the pure leptonic decay such as $B_c \to \tau \bar{\nu}$ is theoretically simple, information from this process is not included in the determination of the CKM matrix due to the lack of accuracy in measurements. Instead, the inclusive and exclusive semileptonic decays are experimentally employed~\cite{10.1093/ptep/ptaa104}.
	Meanwhile, recent experiments observed anomalies in the decays of $\overline{B}\to D^{(*)}\tau \bar{\nu}$ and $B_c\to J/\psi \tau\bar{\nu}$~\cite{PhysRevLett.109.101802,PhysRevD.92.072014,PhysRevD.94.072007,PhysRevLett.118.211801,PhysRevLett.115.111803,PhysRevLett.120.171802,PhysRevLett.120.121801}, which could indicate the existence of physics beyond the standard model. These processes have been studied in non-relativistic quantum chromodynamics (QCD)~\cite{Bell:2005gw,Bell:2008er}, covariant light-front quark model~\cite{Wang:2008xt}, QCD sum rules~\cite{Azizi:2019aaf}, and various other theoretical approaches~\cite{Wang_2013,ISSADYKOV2018178,Colquhoun:2016osw,PhysRevD.80.054016,PhysRevD.68.094020}. These results warrant extra efforts to investigate the process of semileptonic decays of the $B$ and $B_c$ mesons.
	
	The semileptonic decays can be described by a set of scalar functions known as form factors.
	In this work we study the weak decay form factors of $B_c$ based on the basis light-front quantization (BLFQ), a light-front Hamiltonian formalism of quantum field theories~\cite{PhysRevC.81.035205}. 
	However, the physically allowed transferred momentum in the semileptonic decay form factors is timelike, whereas in the light-front approach the traditional choice of the Drell-Yan frame only allows for spacelike momentum transfer.
	Therefore, within such a frame, one needs to either apply analytical continuation or use a factorization approach that introduces extra parameters to reach the timelike region~\cite{PhysRevD.67.113007,PhysRevD.53.1349,PhysRevD.79.054012,PhysRevD.97.054014,KISELEV2000473,Wang_2013,ISSADYKOV2018178,PhysRevD.79.034004,Rui2016,PhysRevD.74.074008}. Meanwhile, since the mass difference between $B_c$ and $ \eta_c\ (J/\psi)$ is rather large, it is technically difficult for some of the approaches to reach the zero-recoil point of $q^2_\text{max} = (M_{B_c} - M_{\eta_c (J/\psi)})^2\approx11 (10)\text{ GeV}^2$~\cite{Colquhoun:2016osw}. In this article we explore frames other than the Drell-Yan frame to gain access to the full physical region of these form factors using light-front quantum field theories. 
	
	Specifically, the weak transition form factors of $B_c$ into either $\eta_c$ or $J/\psi$ are closely related to the matrix elements of the quark current operator. Calculating these matrix elements requires the knowledge of the light-front wave functions (LFWFs) for these mesons. Previously, we have studied the heavy meson systems~\cite{LI2016118,PhysRevD.96.016022,PhysRevD.98.114038} in the framework of BLFQ. Those studies showed success in predicting the mass spectrum and in producing expected LFWFs within the valence Fock sector ($\ket{q\bar{q}}$). After obtaining these LFWFs, one can calculate the observables of interest, such as the form factors of radiative transitions, parton distribution functions, and generalized parton distributions~\cite{PhysRevD.98.034024,PhysRevD.102.014020,PhysRevC.99.035208}. Now we use the obtained LFWFs of $B_c$ and $\eta_c$ ($J/\psi$) to calculate the form factors of weak transitions between them.
	In principle, form factors are Lorentz invariants therefore independent of the reference frames. However we observe frame dependence of form factors that has also been discussed in the literature~\cite{PhysRevD.67.113007,PhysRevD.80.054016}, which is an artifact of Fock space truncation. As more Fock sectors are included, such frame dependence is expected to be reduced~\cite{LI2015278,PhysRevD.94.096008}. In the present work we retain only the leading Fock sector of mesons and seek a frame where errors due to omitting higher Fock sector are minimized. In addition, we investigate how the basis cutoff affects these form factors and their frame dependence.
	
	We organize this paper as follows: in Sec.~\ref{sec_ffonLF} we discuss the semileptonic decay on the light front, and introduce two boost invariants to describe the light-front kinematics. Then we present the calculated results in Sec.~\ref{sec_numres} where we also compare with several other theoretical approaches. Sec.~\ref{sec_summary} contains our conclusion and proposes possible future improvements.

	\section{Semileptonic Decays in Light-Front Kinematics}
	\label{sec_ffonLF}
	\subsection{Form factors and decay width}
	The hadronic matrix element describing the electroweak decays of the lowest $B_c \ (b\bar{c})$ state  is given by
	\begin{equation}
	\mathcal{M}^\mu_h =\mel{P_2, m_j}{V^\mu - A^\mu}{P_1},
	\label{eq_current}
	\end{equation}
	where $P_1$ and $P_2$ are the 4-momentum of the initial and final states, respectively. Here  $m_j$ is the angular momentum magnetic projection of the daughter meson. $V^\mu$ and $A^\mu$ are the vector and axial-vector currents of quark fields, respectively. In particular, we consider the $b \to c$ decay via the emission of a $W^-$ boson. The hadronic matrix element defined by Eq.~\eqref{eq_current} can be parameterized by a set of form factors as functions of the Lorentz invariant $q^2$, where $q^\mu = (P_1-P_2)^\mu$ is the  momentum transfer between the initial and final hadrons.
	We first consider the decay of $B_c$ to an $\eta_c$ plus a pair of leptons. The hadronic matrix elements for the transition between two pseudoscalar (P) mesons are given by~\cite{PhysRevD.37.681}
	\begin{equation}
		\begin{aligned}
			\mel{P_2}{A^\mu}{P_1} &= \mel{P_2}{\bar{c} \gamma^\mu\gamma_5 b}{P_1}  = 0,\\
			\mel{P_2}{V^\mu}{P_1} & = \mel{P_2}{\bar{c} \gamma^\mu b}{P_1} \\
			& = f_+(q^2)P^\mu +f_-(q^2)q^\mu,
		\end{aligned}
		\label{eq_PSdecom}
	\end{equation}
	where $P^\mu =  (P_1+P_2)^\mu$. An alternative to Eq.~\eqref{eq_PSdecom} is another widely used expression for the vector current matrix element in terms of $f_+(q^2)$ and $f_0(q^2)$, which are also known as the transverse and the longitudinal form factors~\cite{Wirbel1985}:
	\begin{equation}
		\begin{aligned}
			\mel{P_2}{V^\mu}{P_1} & = f_+(q^2) \left(P^\mu-\frac{M_1^2-M_2^2}{q^2} q^\mu\right) \\
			& \quad + f_0(q^2) \frac{M_1^2-M_2^2}{q^2}q^\mu.
		\end{aligned}
	\end{equation}
	Here $M_1$ and $M_2$ stand for the masses of the mother meson and the daughter meson, respectively. Consequently, $f_0$ can be written as a linear combination of $f_+$ and $f_-$:
	\begin{equation}
	f_0(q^2) = f_+(q^2) +\frac{q^2}{M_1^2-M_2^2}  f_-(q^2).
	\end{equation}
	The differential decay width for the exclusive process $\text{P} \to \text{P} \ell \bar{\nu}_\ell$ ($\ell = e,\ \mu, \text{ and } \tau$) is~\cite{PhysRevD.59.034001}
	\begin{equation}
		\begin{aligned}
			& \quad \frac{\dd \Gamma(\text{P} \to \text{P} \ell \bar{\nu}_\ell)}{\dd q^2}  = \frac{G^2_\text{F} \abs{V_{cb}}^2}{24\pi^3}  K(q^2) \left(1-\frac{m^2_\ell}{q^2}\right)^2 \\
			& \quad \quad \times \bigg\{ K^2(q^2) \left(1+\frac{m^2_\ell}{2q^2} \right) \abs{f_+(q^2)}^2  \\
			& \quad + M_1^2 \left(1-\frac{M_2^2}{M_1^2} \right)^2 \frac{3 m^2_\ell}{8 q^2} \abs{f_0(q^2)}^2 \bigg\}.
		\end{aligned}
		\label{eq_widPS}
	\end{equation}
	In the expression above we take into account the lepton mass $m_\ell$. ${V_{cb}}$ is the element of the CKM mixing matrix, $G_\text{F}$ is the Fermi coupling constant, and $K(q^2)$ is a kinematic factor defined as
	\begin{equation}
	K(q^2)=\frac{1}{2M_1} \sqrt{(M_1^2+M_2^2-q^2)^2-4M_1^2M_2^2}.
	\end{equation}
	We present results for $f_+(q^2)$ and $f_0(q^2)$ in this work as well as the results for the differential decay width evaluated with Eq.~\eqref{eq_widPS}.  
	
	In the case of a vector (V) final state, both vector and axial-vector current matrices are nonzero:
	\begin{equation}
		\begin{aligned}
			\label{eq_Vdecom}
			& \quad \mel{P_2,m_j}{V^\mu}{P_1} = \mel{P_2,m_j}{\bar{c} \gamma^\mu b}{P_1}   \\
			& = \imag g(q^2) \varepsilon^{\mu \nu\alpha \beta} \epsilon^*_{\nu} P_\alpha q_\beta, \\
			& \quad \mel{P_2,m_j}{A^\mu}{P_1} =\mel{P_2,m_j}{\bar{c} \gamma^\mu\gamma_5 b}{P_1}  \\
			& = f(q^2) \epsilon^{*\mu} + a_+(q^2) (\epsilon ^* \cdot P) P^\mu  \\
			& \quad + a_-(q^2) (\epsilon ^* \cdot P) q^\mu ,
		\end{aligned}
	\end{equation}
	where $\epsilon^* = \epsilon^*(P_2,m_j)$ is the polarization vector of the final meson that satisfies the Lorentz condition $ \epsilon^*\cdot P_2 =0$. These form factors defined in Eq.~\eqref{eq_Vdecom} are related to form factors given by the Bauer-Stech-Wirbel (BSW) convention~\cite{Wirbel1985,Bauer1989}:
		\begin{equation}
		\begin{aligned}
			V(q^2) &= (M_1+M_2) g(q^2), \\
			A_1(q^2) &= \frac{f(q^2)}{M_1+M_2}, \\	
			A_2(q^2) & = -(M_1+M_2) a_+(q^2), \\
			A_0(q^2) & = \frac{1}{2M_2} \big\{ f(q^2) + (M_1^2-M_2^2)a_+(q^2)  \\
			& \quad + q^2 a_-(q^2) \big\}. \\
		\end{aligned}
		\label{eq_BSW2}
	\end{equation}
	The differential decay width characterizing the $\text{P} \to \text{V} \ell \bar{\nu}_\ell$ process can then be expressed with the form factors in the BSW convention as 
	\begin{equation}
		\begin{aligned}
			& \quad \frac{\dd \Gamma (\text{P} \to \text{V} \ell \bar{\nu}_\ell)}{\dd q^2}  \\
			& = \frac{G_F^2 \abs{V_{cb}}^2}{48 \pi^3} K(q^2)\left(1-\frac{m^2_\ell}{q^2}\right)^2 \Bigg \{
			\left(1+\frac{m^2_\ell}{2q^2}\right)  \\
			& \quad \times \bigg[\left(1+\frac{M_2}{M_1}\right)^2 q^2 \abs{A_1(q^2)}^2 + \frac{1}{2M_1^2M_2^2}  \\
			& \quad \quad \times \bigg\vert \frac{1}{2}(M_1^2-M_2^2-q^2)(M_1+M_2)A_1(q^2)  \\
			& \quad \quad \quad - \frac{2K^2(q^2) M_1^2}{M_1+M_2}A_2(q^2) \bigg\vert ^2  \\
			& \quad \quad + \frac{4  q^2 K^2(q^2) }{(M_1+M_2)^2} \abs{V(q^2)}^2\bigg]  \\
			& \quad + \frac{3m^2_\ell}{q^2} K^2(q^2)  \abs{A_0(q^2)}^2
			\Bigg\}.
		\end{aligned}
	\end{equation}
	The various form factors appearing  in Eqs.~\eqref{eq_PSdecom} and~\eqref{eq_Vdecom} will be defined in terms of LFWFs in Sec.~\ref{sec_numres}.
	
	\subsection{Light-front kinematics}
	
\begin{figure*}
	\hspace{-1cm}
	\centering
	\includegraphics[scale=0.48]{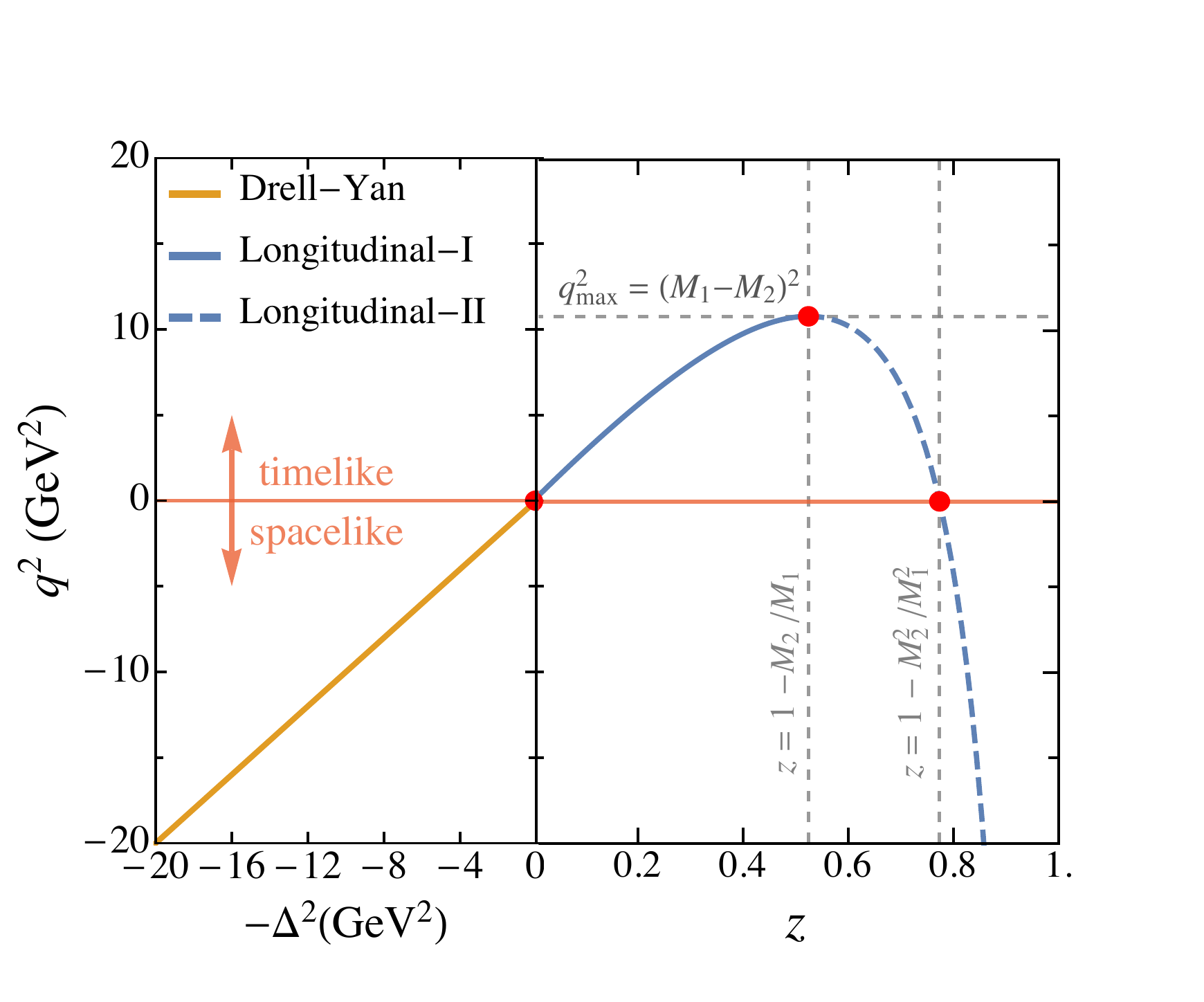}
	\hspace{.1cm}
	\includegraphics[scale=0.54]{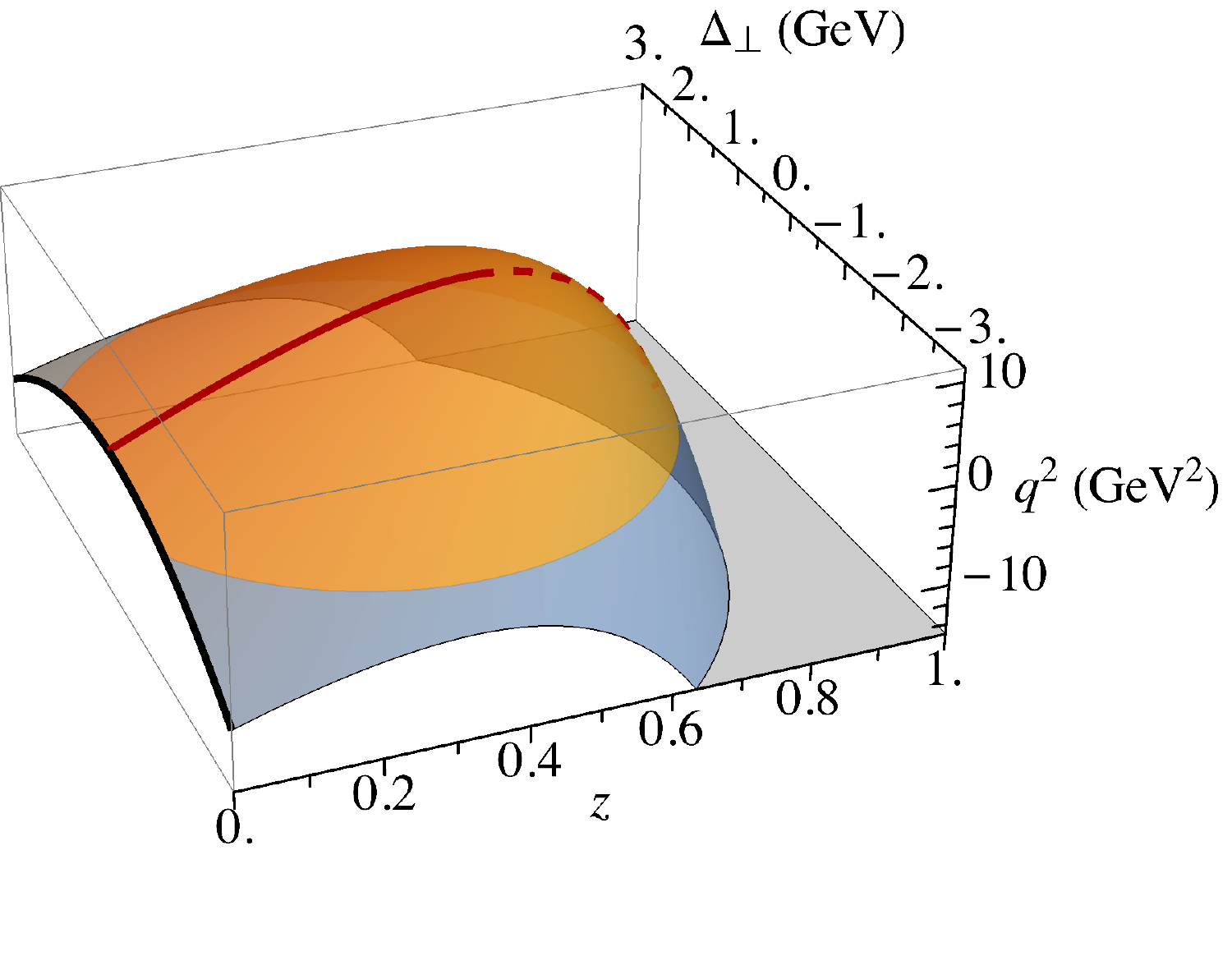}
	\caption{\textbf{Left panel:} The Lorentz invariant $q^2$ as a function of $z$ or $\Delta_\perp^2$ in two special frames. In the Drell-Yan frame ($z=0$), momentum transfer squared is given by $q^2 = -\Delta^2_\bot$. In the longitudinal frame ($\Delta_\perp$=0), $q^2 = z[M_1^2-M_2^2/(1-z)]$. Note that in this 2-dimensional (2D) plot we put the two regions with different frames together only for the sake of visualization, they have different variables and scales on the horizontal axes. \\
		\textbf{Right panel:} The 3-dimensional plot of $q^2$ in terms of $z$ and $\Delta_\bot$. All $q^2$ values in the physically allowed region are situated on the convex surface and correspond to a pair of $(z, \Delta_\bot)$. The yellow area shows the timelike region, while the blue area is spacelike. The black curve corresponds to the Drell-Yan frame, the solid red curve represents the longitudinal-\textrm{I} frame, and the dashed red curve (which drops out of sight over the peak of the convex surface) traces the longitudinal-\textrm{II} frame.}
	\label{fig_frame}
	\end{figure*}

	To describe the kinematics of the decay process in the light-front coordinates where the transferred 4-momentum is defined as $q^\mu = (q^+,\vec{q}_\perp,q^-)=(q^0+q^3,q^1,q^2,q^0-q^3)$, we introduce two boost invariants $z$ and $\vec{\Delta}_\perp$~\cite{PhysRevD.97.054034}. 
	Specifically, $z$ is the relative momentum transfer in the longitudinal direction, and is limited to the kinematical region $0 \le z < 1$ in the valence Fock sector. Meanwhile, $\vec{\Delta}_\bot$ is a combination of the momentum transfer in the transverse direction and the relative longitudinal momentum transfer:
	\begin{equation}
	z=\frac{q^+}{P_1^+}, \qquad \vec{\Delta}_\bot = \vec{q}_\bot - z \vec{P}_{1 \bot}.
	\end{equation}
	Therefore the Lorentz invariant $q^2=q^\mu q_\mu$ can be written in terms of $z$ and $\vec{\Delta}_\perp$ as
	\begin{equation}
	q^2 = z\left(M_1^2 - \frac{M_2^2}{1-z}\right) - \frac{\Delta^2_\bot}{1-z}.
	\label{eq_q2}
	\end{equation}
	Note that in Eq.~\eqref{eq_q2} $q^2$ is not a monotonic function in $z$. Therefore for a given fixed $q^2$ there are different $(z, \vec{\Delta}_\bot)$ combinations defining different frames illustrated in FIG.~\ref{fig_frame}. Among them, the following special frames are of particular interest:
	\begin{itemize}
		\item Drell-Yan frame: $q^+=0 \ (z =0)$. This frame is widely adopted in light-front dynamics, especially working in combination with the ``good current'' $J^+$, which usually yields a concise setup for the calculation. However, by choosing the Drell-Yan frame, one can only access the spacelike region since $q^2=-\Delta^2_\perp \le0$. Recall that in the case of the semileptonic decay of hadrons, the physically allowed region of $q^2$ is timelike. 
		It is therefore common practice to either apply analytic continuation by replacing $ \vec{q}_\perp$ with $\imag \vec{q}_\perp$ \cite{PhysRevD.67.113007,PhysRevD.53.1349} or to use factorization \cite{Wang_2013,ISSADYKOV2018178} in order to access the timelike region.
		\item Longitudinal frame: $\vec{\Delta}_\bot = 0$.  This frame covers both timelike and spacelike regions. 
		Furthermore, it is the only frame that grants access to the zero recoil point, i.e. $q^2_\text{max}=(M_1-M_2)^2$ (see FIG.~\ref{fig_frame}).  Only at this point is $z = 1 - M_2/M_1$ unique. Unlike the Drell-Yan frame where $q^2=-\Delta^2_\perp$, the longitudinal frame has two branches that indicate the existence of two $z$ values contributing to the same $q^2$ (except $q^2_\text{max}$) in the timelike region: ${z=[M_1^2-M_2^2+q^2} \pm {\sqrt{(M_1^2-M_2^2+q^2)^2-4M_1^2q^2}]/(2M_1^2)}$. Thus we treat them as two different frames. The first branch with $0<z<1-M_2/M_1$ is connected to the Drell-Yan frame at $q^2=0$. We refer this branch as the longitudinal-\textrm{I} frame, through which one can only access the timelike region.
		The other branch, referred to as the longitudinal-\textrm{II} frame, starts at the limit $q^2_\text{max}$ for $z = 1 - M_2/M_1$, extending towards spacelike infinity for $z \to 1$. That is, it covers the entire spacelike region, as well as the physical timelike region.
	\end{itemize}
	Aside from these special frames, all other combinations of $z$ and $\Delta_\bot$ on the convex surface in the right panel of FIG.~\ref{fig_frame} constitute the general frames.
		
\begin{figure}
	\centering
	\begin{subfigure}{0.48\textwidth}
		\includegraphics[scale=0.65]{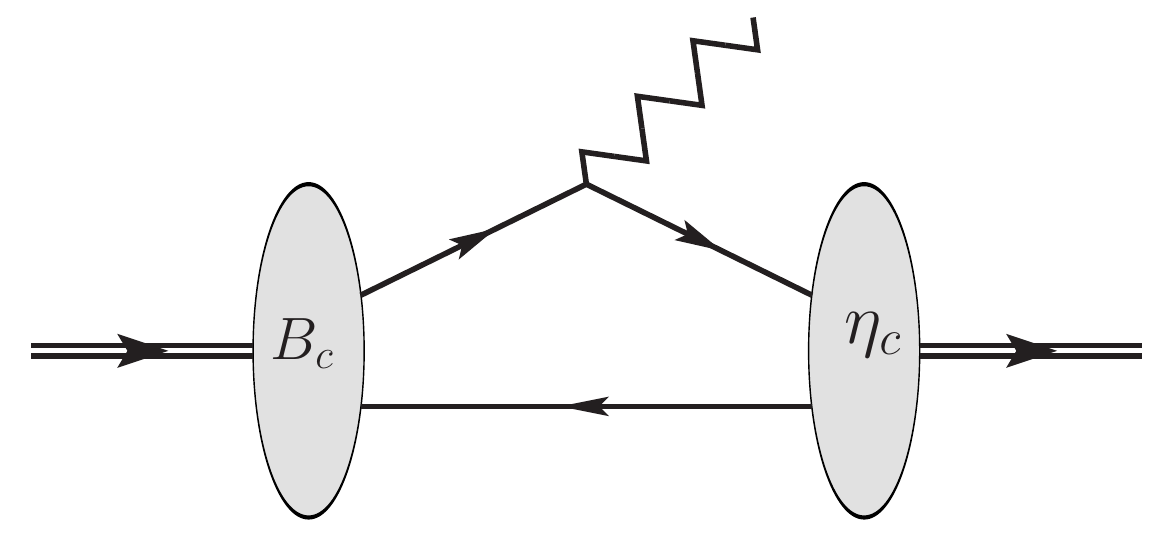}
		\caption{conserving particle number: 2 $\to$ 2}
		\label{fig_decay1}
	\end{subfigure}
	\hspace{.2cm}
	\begin{subfigure}{0.48\textwidth}
		\includegraphics[scale=0.65]{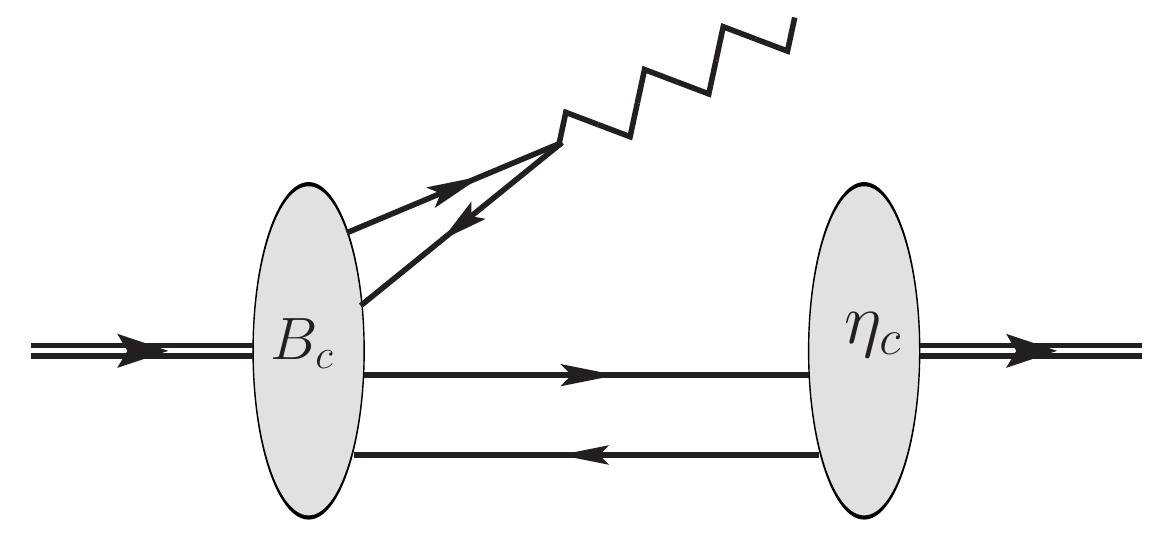}
		\caption{changing particle number: 4 $\to$ 2}
		\label{fig_decay2}
	\end{subfigure}
	\caption{Diagrams of two dominant contributions to the transition $B_c \to \eta_c$ in the light-front time order (left to right).}
	\label{fig_decaydiagram}
	\end{figure}

	Recall that the LFWFs we employ from Refs.~\cite{PhysRevD.96.016022,PhysRevD.98.114038} are constrained within the valence Fock sector $\ket{q\bar{q}}$. Thus we study the decay process without contributions due to particle annihilation. This process is referred to as the leading-order Feynman diagram shown in FIG.~\ref{fig_decay1}.
	Furthermore, the contribution from the particle-number-changing diagrams, which involves Fock sectors higher than the valence such as FIG.~\ref{fig_decay2}, are not yet available in BLFQ~\cite{PhysRevD.97.054034,PhysRevD.100.036006}. In the literature~\cite{PhysRevD.8.4574,PhysRevD.22.2236,BRODSKY1999239}, the Drell-Yan frame together with the good current is adopted since it has the advantage of suppressing the vacuum pair production/annihilation including the case in FIG.~\ref{fig_decay2}; however, this frame has the limitation that it only allows for spacelike $q^2$ as stated before.
	In the following section, we will discuss the variations in the form factors by considering all of the frames specified in right panel of FIG.~\ref{fig_frame}.  We expect that some of these variations are due to the Fock space truncation in our model but incorporating higher Fock spaces explicitly is beyond the scope of the present work.

	\section{\label{sec_numres}Numerical Results}
	
	\subsection{The decay process $B_c$ $\to$ $\eta_c \ell \bar{\nu}_\ell$}
	In BLFQ, we employ an effective Hamiltonian which consists of a 2D harmonic oscillator (HO) transverse confinement, a longitudinal confinement, and an effective one-gluon-exchange potential~\cite{PhysRevD.96.016022,PhysRevD.98.114038}.  In the basis representation, the 2D HO function with parameter $\kappa$ is adopted for the transverse direction, while we employ the Jacobi polynomials in the longitudinal direction (see Appendix~\ref{appxA} and Refs.~\cite{PhysRevD.96.016022,PhysRevD.98.114038} for details).
	The basis cutoff $N_\text{max}$ acts implicitly as the infrared (IR) and ultra-violet (UV) regulators for the LFWFs in the transverse direction, with an IR cutoff $\lambda_\text{IR} \approx \kappa/\sqrt{N_\text{max}}$ and a UV cutoff  $\lambda_\text{UV} \approx  \kappa \sqrt{N_\text{max}}$. The basis cutoff $L_\text{max}$ controls the numerical resolution and regulates the longitudinal direction. For convenience, and in accordance with previous work, $L_\text{max}$ and $N_\text{max}$ are taken to be equal for the wave functions we employ in this paper. The hadronic matrix elements, expressed as the overlaps of LFWFs, depend on these cutoffs as we will discuss below.~\footnote{In this article, we use ``basis cutoff'' or ``basis dependence'' to indicate the dependence on all the model parameters $m_q$, $\kappa$, and $N_\text{max}=L_\text{max}$. Details of parameters can be found in Refs.~\cite{PhysRevD.96.016022,PhysRevD.98.114038}.}
	
	In Eq.~\eqref{eq_PSdecom}, the hadronic matrix element of the vector current coupled to two pseudoscalar mesons is decomposed in terms of two Lorentz scalar functions, i.e. the transition form factors $f_\pm(q^2)$. In perturbative QCD, this hadronic matrix element can be factorized in terms of the meson distribution amplitudes. The form factors are then expressed as convolutions of these amplitudes with hard
	scattering kernels~\cite{PhysRevD.65.014007,WEI2002263}. 
	However, instead of using these distribution amplitudes, we calculate the hadronic matrix elements using the LFWFs of mesons directly without factorization. To construct equations for the transition form factors in Eq.~\eqref{eq_PSdecom}, we employ two current components, $\mu = +$ and $\mu = R$. The $R$ component of a vector is defined by $x^R \triangleq x^1 + \imag x^2$, with the $L$ component being its conjugate. We do not use the $\mu=-$ component as it violates charge conservation within our truncated Fock sector.
	The matrix elements $\mathcal{M}^{+(R)} \triangleq \mel{P_2}{\bar{c}\gamma^{+(R)}b}{P_1}$ can then be calculated through the overlap integral of the meson wave functions according to 
	\begin{equation}
		\begin{aligned}
			& \quad \mathcal{M}^\mu  \\
			& = \sum_{s\bar{s}}\int_z^1 \frac{\dd x}{2x(1-x)} \int \frac{\dd \vec{k}_\perp}{(2\pi)^3} \sum_{s'} \frac{1-z}{x-z} \bar{u}_{s'}  \left(   p'\right)  \\
			& \quad \times\gamma^\mu u_s(p) \psi^*_{s'\bar{s}/\eta_c} 
			\left(\frac{x-z}{1-z}, \vec{k}_\perp -\frac{1-x}{1-z}\vec{\Delta}_\perp \right)  \\
			&\quad \times \psi_{s\bar{s}/B_c} (x, \vec{k}_\perp),
		\end{aligned}
		\label{eq_matrix}
	\end{equation}
	where $p = {\left(x P_1^+, \vec{k}_\perp + x \vec{P}_{1\perp}\right) }$ and $p'= {\left((x-z)P^+_1 ,\vec{k}_\perp +(x-z) \vec{P}_{1\perp} -\vec{\Delta}_\perp\right) }$ are the 3-momentum of the initial quark ($b$) and final antiquark ($\bar{c}$) associated with the spinors $u_s$ and $\bar{u}_{s'}$. Details of the LFWFs $\psi(x,\vec{k}_\perp)$ expanded within the BLFQ basis representation are given in Appendix \ref{appxA}. Meanwhile, tables of the spinor matrix components for our applications are provided in Appendix \ref{appxB}.
	With Eq.~\eqref{eq_matrix}, the two form factors in Eq.~\eqref{eq_PSdecom} can be written in terms of the matrix elements and the two boost invariants $z$ and $\Delta_\perp$:
	\begin{equation}
		\begin{aligned}
			& f_+(q^2) =\frac{(\Delta^R+z P_1^R)\mathcal{M}^+ - z P_1^+ \mathcal{M}^R }{2 \Delta^R P_1^+ },\\
			& \quad f_-(q^2)  \\
			&= \frac{\left[\Delta^R-(2-z)P_1^R\right]\mathcal{M}^+ +(2-z)P_1^+ \mathcal{M}^R}{2 \Delta^R P_1^+ }.\\
		\end{aligned}
	\end{equation}
	
\begin{figure*}
		\hspace{-1cm}
		\begin{subfigure}[t]{0.48\textwidth}
			\includegraphics[scale=0.65]{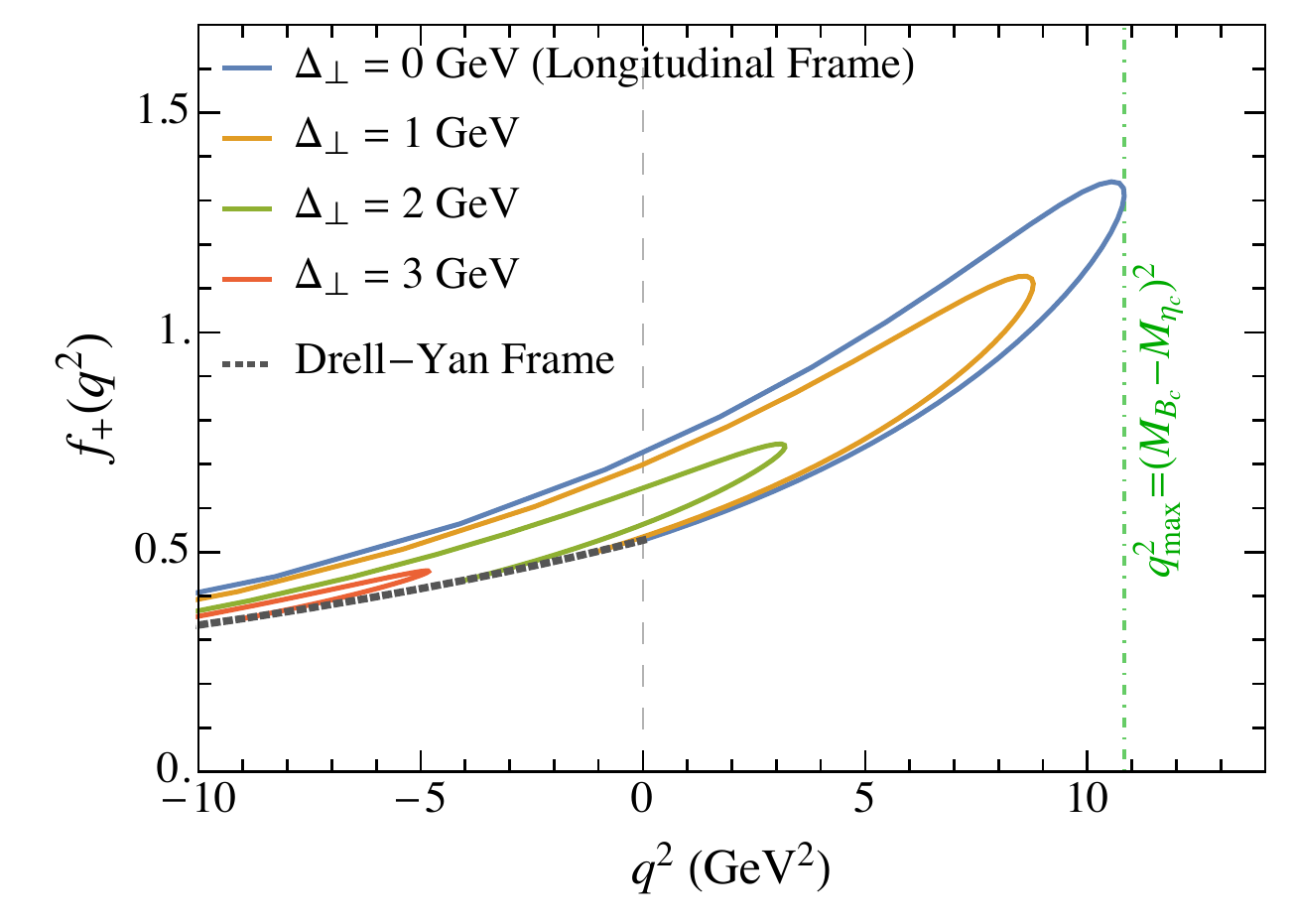}
		\end{subfigure}
		\hspace{1cm}
		\begin{subfigure}[t]{0.48\textwidth}
			\includegraphics[scale=0.65]{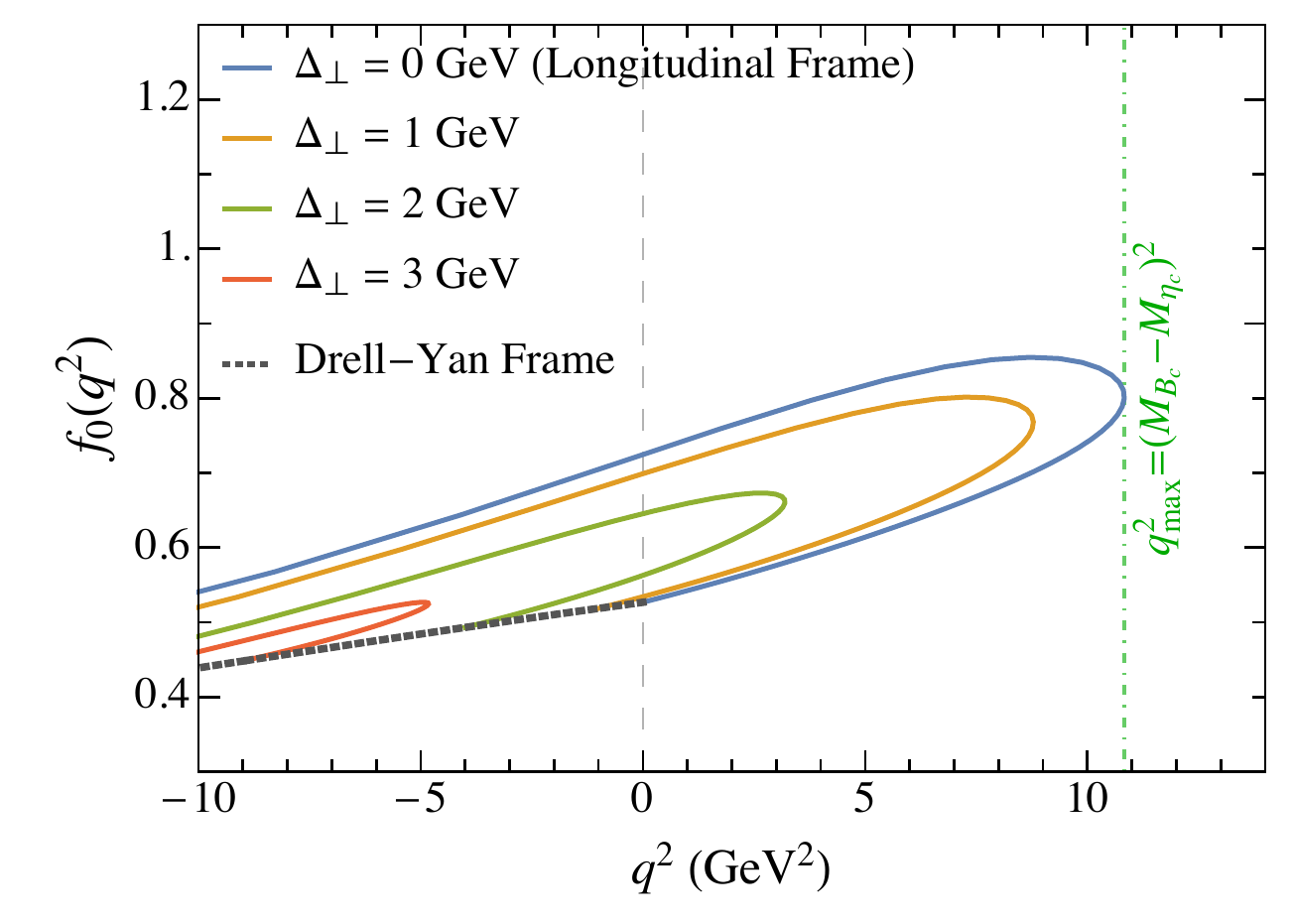}
		\end{subfigure}
		\caption{The frame dependence of the form factors for $B_c \to \eta_c  \ell \nu_\ell$ decay at the basis size $N_\text{max}=L_\text{max}=8$. The black dots indicate the form factors obtained with the Drell-Yan frame, the solid blue curve is from the longitudinal frame, and the other curves are obtained with both nonzero $z$ and $\Delta_\perp$ which we call the general frames. For $f_+(q^2)$ and $f_0(q^2)$, the two special frames form a boundary enclosing form factors obtained from the general frames.}
		\label{fig_framePS}
	\end{figure*}

	We present the results of $f_+(q^2)$ and $f_0(q^2)$ obtained with different frames in FIG.~\ref{fig_framePS}, where we sample the $q^2$ with respect to multiple $\Delta_\bot$ and $z$ pairs. The difference of the form factors between two different frames is referred to as the frame dependence. The constraint $z = 0$ selects the Drell-Yan frame which only supports $q^2 \le 0$. It is connected to the longitudinal-\textrm{I} frame (lower branch of the longitudinal frame curve) at $q^2 = 0$. However, there is a kink at this connecting point which indicates that the derivative with respect to $q^2$ is not continuous at the boundary of the two frames. Therefore, one has to be careful when applying the analytical continuation of the form factor from the spacelike region to the timelike region.
	In the timelike region, form factors with the longitudinal frames I and II correspond to the lower and upper limits of $f_{+,0}(q^2)$ in the figure, whereas the form factors calculated in general frames with nonzero $\Delta_\perp$ fall  between these two curves.  Similarly, for small spacelike momenta, the longitudinal-II and Drell-Yan results for the form factors correspond to the upper and lower limits of results obtained in the general frames, but as one moves further into the spacelike region, results in the general frames may exceed these two boundaries. However, for the weak decay processes considered in this article we are mainly interested in the physical timelike region. 
	
\begin{figure*}
		\hspace{-1cm}
		\begin{subfigure}[t]{0.48\textwidth}
			\includegraphics[scale=0.65]{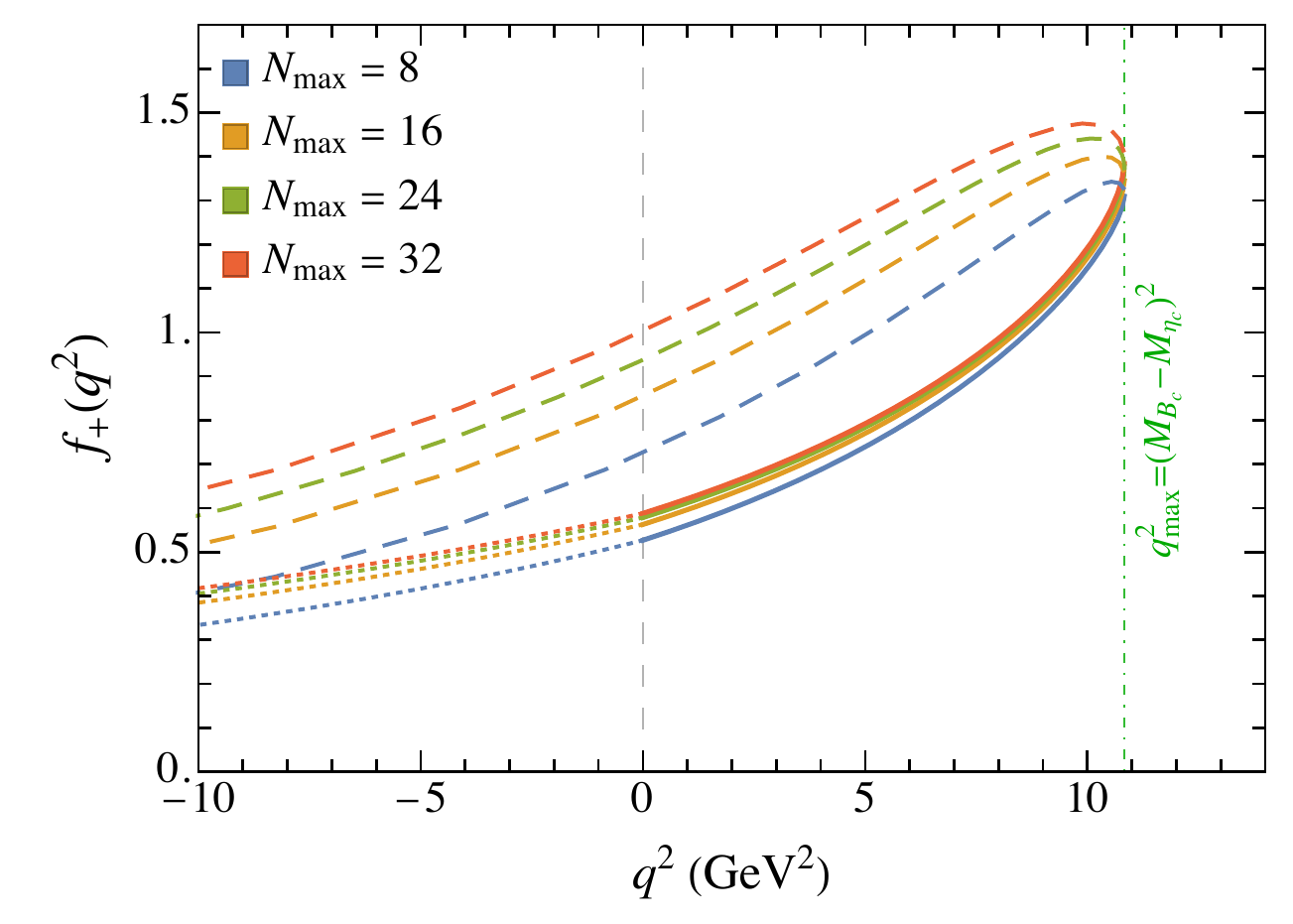}
		\end{subfigure}
		\hspace{1cm}
		\begin{subfigure}[t]{0.48\textwidth}
			\includegraphics[scale=0.65]{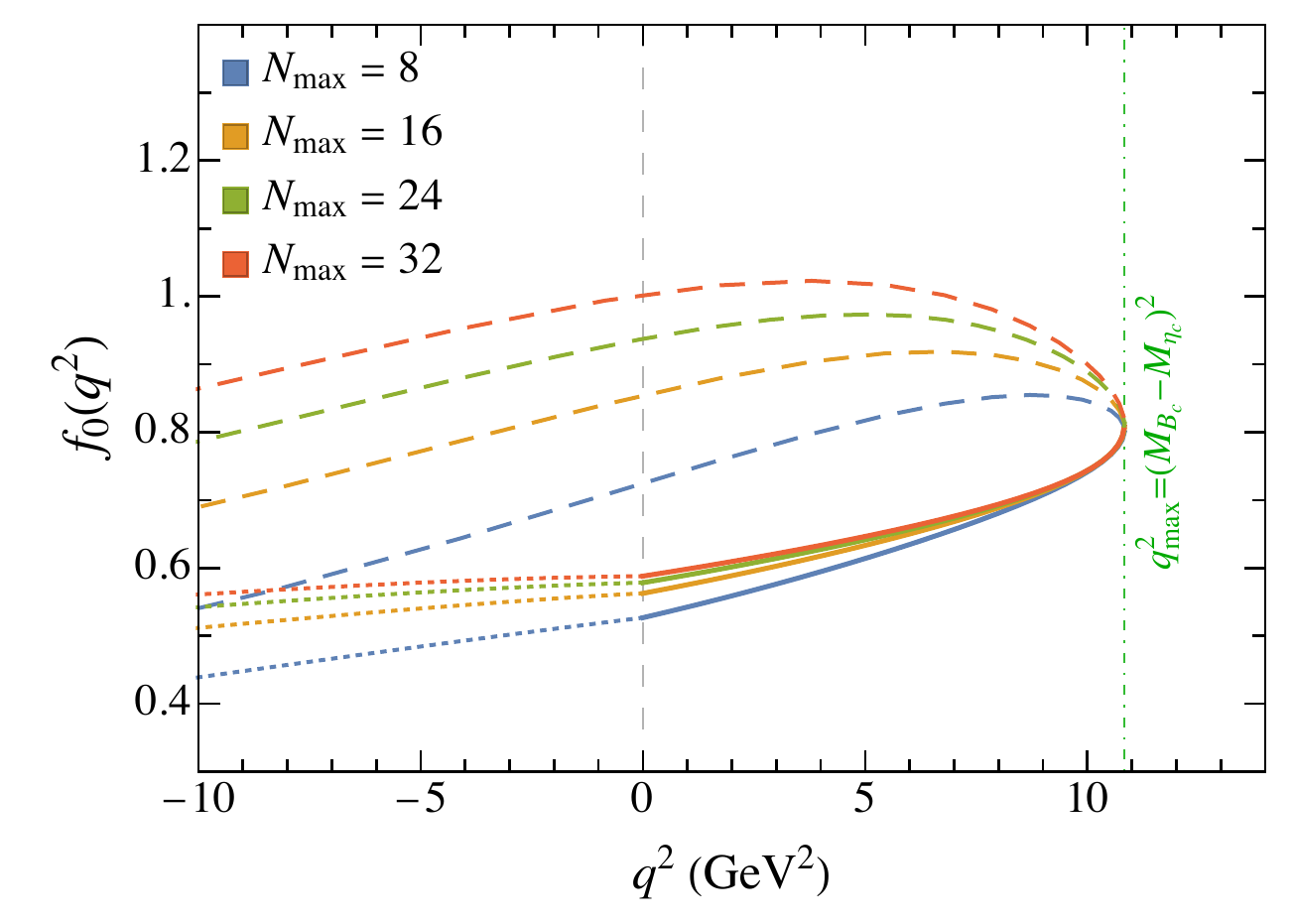}
		\end{subfigure}
		\caption{The dependence of the form factors on the basis cutoff. Different colors are used to distinguish the basis cutoffs, i.e. blue, yellow, green, and red are representing $N_\text{max} = 8, \ 16, \ 24,\ 32$, repetitively (same for FIG.~\ref{fig_widthPS},~\ref{fig_basis_ffV}, and~\ref{fig_widthV} below). Meanwhile, we present the form factors calculated in the Drell-Yan frame with the dots; results in the longitudinal-\textrm{I} and longitudinal-\textrm{II} frames are shown with solid and dashed curves, respectively. }
		\label{fig_fpfm}
	\end{figure*}

	In order to gain a clear impression on how basis truncation affects the form factors, we include the special frames with basis cutoffs $N_\text{max}=L_\text{max}=8, \ 16, \ 24,$ and $32$ in FIG.~\ref{fig_fpfm}. 
	We notice that the dependence on $N_\text{max}$ is decreasing as $N_\text{max}$ increases within each frame for $f_+(q^2)$ and $f_0(q^2)$. Note that the dependence on $N_\text{max}$ of form factors is almost negligible at $q^2_\text{max}$,
	but the frame dependence becomes stronger with a larger basis size arising primarily from a more significant increase in the results of the longitudinal-\textrm{II} branch in comparison with the Drell-Yan and longitudinal-\textrm{I} branches. 	
	This overall tendency appears to be counterintuitive since larger basis size usually reduces the frame dependence in several other observables, such as the decay constant, the elastic form factor, and radiative transition form factors~\cite{PhysRevD.98.114038,Tang:2019gvn,PhysRevD.97.054034,PhysRevD.98.034024,PhysRevD.100.036006}. We recall that the role of the omitted particle-number-changing diagram is suppressed by keeping to a value of $z$ as low as possible.  

	In contrast, the longitudinal-II frame with a larger $z$ increases the weight of the longitudinal tails of LFWFs in the $x$-integral of Eq.~\eqref{eq_matrix}, and results in a larger dependence of the basis cutoffs~\cite{Maris:2020wew}.
	It is therefore appealing that the smaller sensitivity to basis space cutoff of the Drell-Yan and longitudinal-\textrm{I} frames compared to others suggests these two frames are preferred for calculations with the currently available LFWFs.
	
	In TABLE~\ref{tb1}, we list the numerical values of the form factors at the kinematical limits, namely $q^2 = 0$ and $q^2_\text{max}$ at $N_\text{max}=L_\text{max}=32$. We also provide results from other approaches for comparison. In general, our result with longitudinal-\textrm{I} frame shows better agreement with other approaches.  This is encouraging in light of our discussion above concerning the minimization of effects of the neglected particle-number-changing term with the longitudinal-\textrm{I} frame.
	
\begin{figure}
		\begin{subfigure}[t]{0.5\textwidth}
			\includegraphics[scale=0.48]{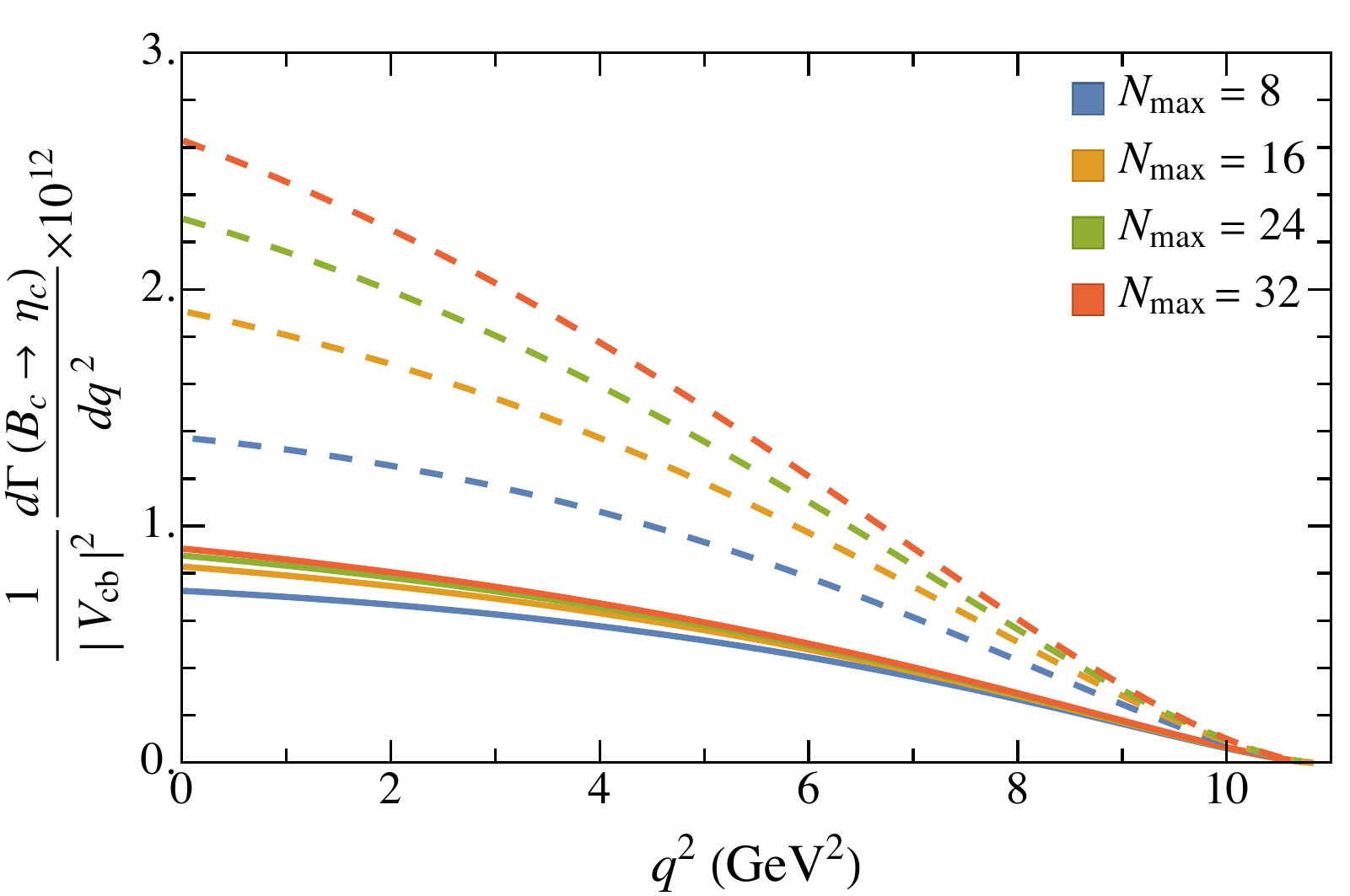}
		\end{subfigure}
		\caption{The differential decay width of semileptonic decay of $B_c \to \eta_c e \bar{\nu}$. Results are presented with longitudinal-\textrm{I} (solid) and longitudinal-\textrm{II} (dashed) frames at different basis cutoffs.}
		\label{fig_widthPS}
	\end{figure}

	In FIG.~\ref{fig_widthPS}, we provide our result for the differential decay width based on Eq.~\eqref{eq_widPS}, specifically for the $B_c\to \eta_c e \bar{\nu}$ channel. Values of the lepton mass and the Fermi constant are taken from the Particle Data Group (PDG)~\cite{10.1093/ptep/ptaa104}:
	\begin{equation}
		\begin{aligned}
			& m_e =0.5109989461 \text{ MeV;}  \\
			& G_\text{F} = 1.166 378 7 \times 10^{-5} \text{ GeV}^{-2}.
		\end{aligned}
	\end{equation}
	The decay width shows similar trends as those found for the form factors. To be specific, results with the longitudinal-\textrm{I} frame are less sensitive to basis sizes than those with longitudinal-\textrm{II}. In addition, the difference between two frames is most significant at low $q^2$. We therefore adopt the longitudinal-\textrm{I} frame as our preferred reference frame for the application of valence LFWFs to semileptonic decay processes. 
			
	\begin{table*}
		\centering 
		\begin{tabular}{c|cccccccccc}
			\toprule
			&  \hspace{0.1cm}  	\multirow{2}{*}{ $f_+(0) = f_0(0)$ }  \hspace{0.2cm} &  \hspace{0.3cm}  	\multirow{2}{*}{ $f_+(q^2_\text{max})$} \hspace{0.3cm}  &   \hspace{0.3cm} 	\multirow{2}{*}{ $f_0(q^2_\text{max})$}  \hspace{0.3cm} \\
			\\
			\midrule
			BLFQ - 1 & $0.588(19)	$	& \multirow{2} {*}{$1.391(35)$}&  \multirow{2} {*}{$0.811(3)$} \\
			BLFQ - 2 & $1.003(130)	$	&  & \\
			pQCD~\cite{Wang_2013} & 0.48 &1.03 &0.78 &  \\
			CCQM~\cite{ISSADYKOV2018178} & 0.75  & 1.13 & 0.92 \\
			LFQM~\cite{PhysRevD.80.054016}& $0.482\ [0.546]$	& $1.084\ [1.035]$  & $0.876\ [0.872]$\\
			RQM~\cite{PhysRevD.68.094020} & 0.47& 1.07 &0.92 \\
			Lattice QCD~\cite{Colquhoun:2016osw} & 0.59& &  \\    
			\bottomrule
		\end{tabular}
		\caption{Form factors calculated in this work within different frames (first two rows of the table) and with other methods. BLFQ - 1 and BLFQ - 2 at $q^2=0$ correspond to longitudinal-\textrm{I} (our preferred frame as discussed in the text) and longitudinal-\textrm{II} frames, respectively. The BLFQ results are quoted using the LFWFs at the basis cutoff $N_\text{max}=L_\text{max}=32$, while the uncertainties are quoted as $\varepsilon_f = 2 \abs{f_{N_\text{max}=32} - f_{N_\text{max}=24}}$ to show the basis sensitivity. Other methods listed in the table include perturbative QCD (pQCD), covariant confined quark model (CCQM), relativistic quark model (RQM), light-front quark model (LFQM), and Lattice QCD. For the LFQM, we quote the results with both linear potential and harmonic oscillator potential (the latter in the square bracket).}
		\label{tb1}
	\end{table*}

	\subsection{The decay process $B_c$ $\to$ $J/\psi \ell \bar{\nu}_\ell$}
	
	The hadronic matrix elements describing the semileptonic decay from $B_c$ to the vector meson $J/\psi$ have more intricate structures than those with a pseudoscalar final state. The vector current matrix element in Eq.~\eqref{eq_Vdecom} takes a form similar to the radiative decay matrix element between pseudoscalar and vector states~\cite{PhysRevD.73.074507}. Based on a previous study~\cite{PhysRevD.98.034024}, a combination of the current component $\mu=R$ and the magnetic projection $m_j=0$ of the LFWF for the final state meson is favored in calculating the form factor $g(q^2)$ on the light front. In addition, this choice employs the dominant spin component of the LFWFs and ties in with the non-relativistic limit of the heavy systems. Consequently, we use the $R$ component of the hadronic matrix element $\mathcal{V}^R_0 \triangleq \mel{P_2,m_j=0}{\bar{c}\gamma^R b}{P_1}$ for the form factor $g(q^2)$:
	\begin{equation}
	g(q^2)=\frac{\imag}{2M_2 } \frac{1-z}{\Delta^R} \mathcal{V}^R_0 ,
	\end{equation}
	where the matrix element $\mathcal{V}^\mu_{m_j}$ is defined in terms of LFWFs as
	\begin{equation}
		\begin{aligned}
			& \quad \mathcal{V}^\mu_{m_j}  \\
			& = \sum_{s\bar{s}}\int_z^1 \frac{\dd x}{2x(1-x)} \int \frac{\dd \vec{k}_\perp}{(2\pi)^3} \sum_{s'} \frac{1-z}{x-z} \bar{u}_{s'}\left(   p'\right)  \\
			& \quad \times \gamma^\mu u_s(p) \psi^{*(m_j)}_{s'\bar{s}/J/\psi} 
			\left(\frac{x-z}{1-z}, \vec{k}_\perp -\frac{1-x}{1-z}\vec{\Delta}_\perp \right)  \\
			& \quad \times \psi_{s\bar{s}/B_c} (x, \vec{k}_\perp).
		\end{aligned}
	\end{equation}
	For the other three form factors corresponding to the axial current matrix element, we employ $\mathcal{A}^+_0$, $\mathcal{A}^+_1$, and $\mathcal{A}^L_1$ for the calculation, where 
	\begin{equation}
		\begin{aligned}
			& \quad \mathcal{A}^\mu_{m_j} \triangleq \mel{P_2, m_j}{\bar{c}\gamma^\mu\gamma_5b}{P_1}  \\
			& = \sum_{s\bar{s}}\int_z^1 \frac{\dd x}{2x(1-x)} \int \frac{\dd \vec{k}_\perp}{(2\pi)^3} \sum_{s'} \frac{1-z}{x-z} \bar{u}_{s'}\left(   p'\right)  \\
			& \quad \times \gamma^\mu \gamma_5 u_s(p) \psi^{*(m_j)}_{s'\bar{s}/J/\psi} 
			\left(\frac{x-z}{1-z}, \vec{k}_\perp -\frac{1-x}{1-z}\vec{\Delta}_\perp \right)  \\
			& \quad \times \psi_{s\bar{s}/B_c} (x, \vec{k}_\perp).
		\end{aligned}
	\end{equation}
	Again, we avoid using the ``bad current'' ($\mu=-$) to calculate form factors in the vector final state decays.
	Then those form factors can be expressed as follows:
	\begin{equation}
		\begin{aligned}
			& f(q^2) = \frac{M_2}{(1-z)P_1^+}\mathcal{A}^+_0  \\
			& \quad \hspace{1cm}+ \frac{\Delta^2_\perp - M_2^2+(1-z)^2M_1^2}{\sqrt{2}(1-z)P_1^+\Delta^L} \mathcal{A}^+_1, \\
			& a_+(q^2) =(1-z) \frac{zP_1^+ \mathcal{A}^L_1 - \left(zP_1^L+\Delta^L\right)\mathcal{A}^+_1}{\sqrt{2}\left(\Delta^L\right)^2P_1^+}, \\
			& a_-(q^2) = (1-z) \\
			& \quad \times \frac{(z-2)P_1^+\mathcal{A}^L_1- \left[(z-2)P^L_1+\Delta^L\right]\mathcal{A}^+_1}{\sqrt{2}\left(\Delta^L\right)^2P_1^+}.
			\label{eq_ffvec}
		\end{aligned}
	\end{equation}
	We use the hadronic matrix elements to calculate the form factors by the relations above, then convert them into the BSW conventions according to Eq.~\eqref{eq_BSW2}.
	Note that the form factor $f(q^2)$ depends on both the $\mathcal{A}^+_0$ and  $\mathcal{A}^+_1$ matrix elements. That is, $f(q^2)$ depends on the LFWFs of $J/\psi$ with both $m_j=0$ and $m_j=1$. However in BLFQ we compute the LFWFs with fixed $m_j$ independently, leaving the relative phase undetermined between any two states with different $m_j$. Subsequently, we determine the relative sign between $\psi^{(m_j=0)}_{J/\psi}$ and $\psi^{(m_j=1)}_{J/\psi}$ by comparing the non-relativistic component of the LFWFs, and ensure that the light-front eigenstates approximately satisfy $\mathcal{J}_+ \ket{m_j=0} = C \ket{m_j=1}$, where $C$ is a positive coefficient corresponding to the ladder operator $\mathcal{J}_+$ on the light front.
	
\begin{figure*}
		\hspace{-1cm}
		\begin{subfigure}[t]{0.48\textwidth}
			\includegraphics[scale=0.65]{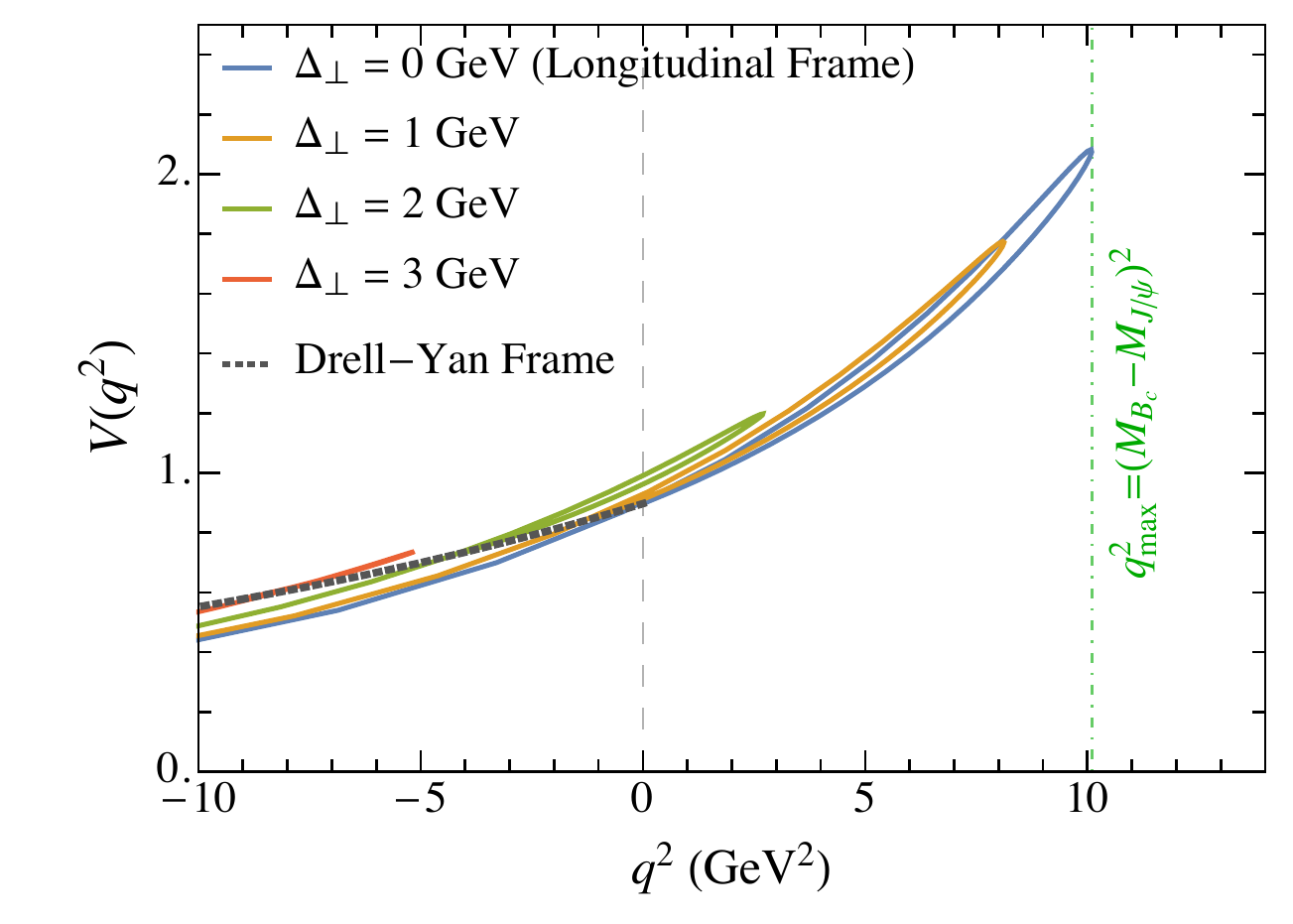}
		\end{subfigure}
		\hspace{1cm}
		\begin{subfigure}[t]{0.48\textwidth}
			\includegraphics[scale=0.65]{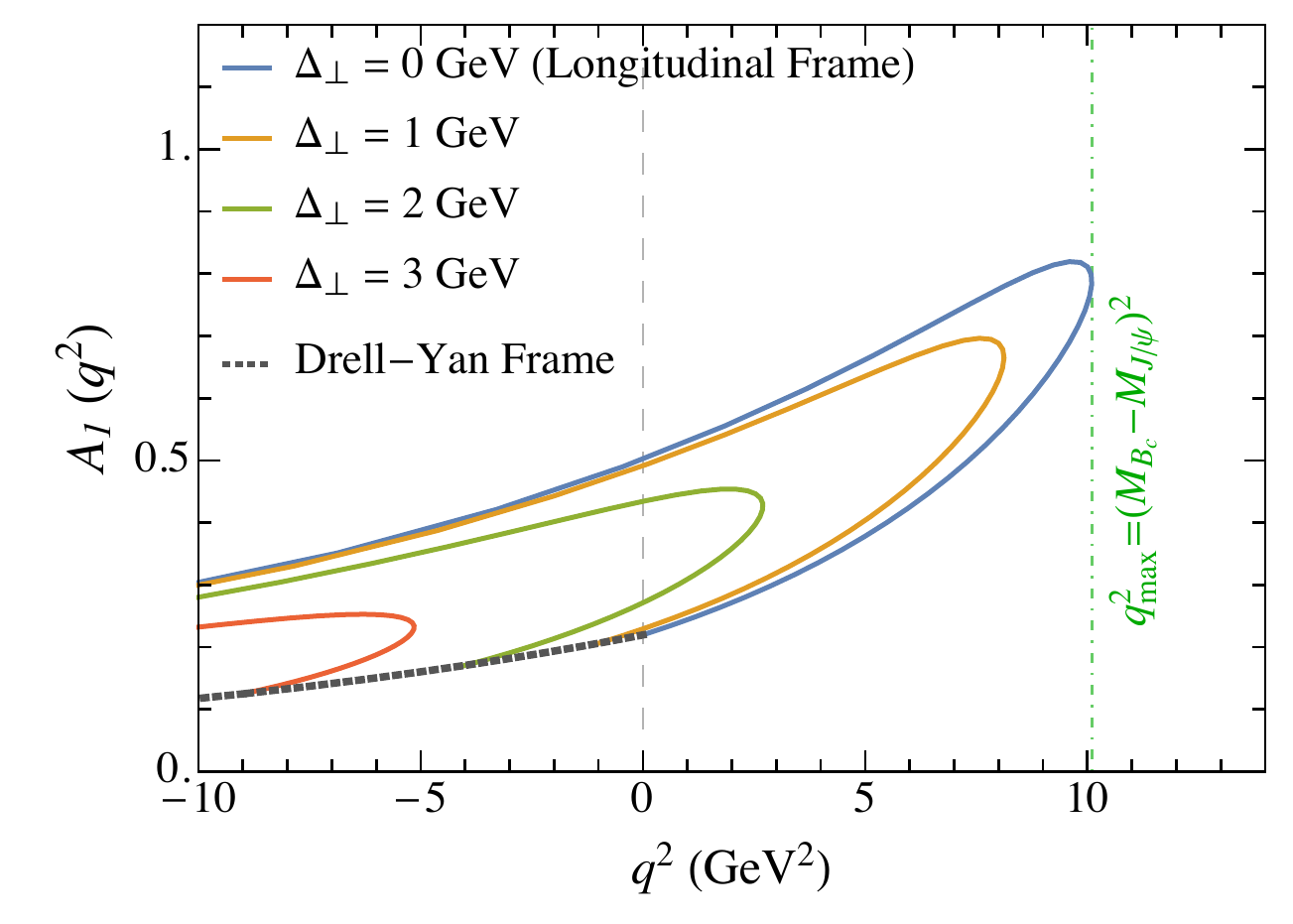}
		\end{subfigure}\\
		\hspace{-1cm}
		\begin{subfigure}[t]{0.48\textwidth}
			\includegraphics[scale=0.65]{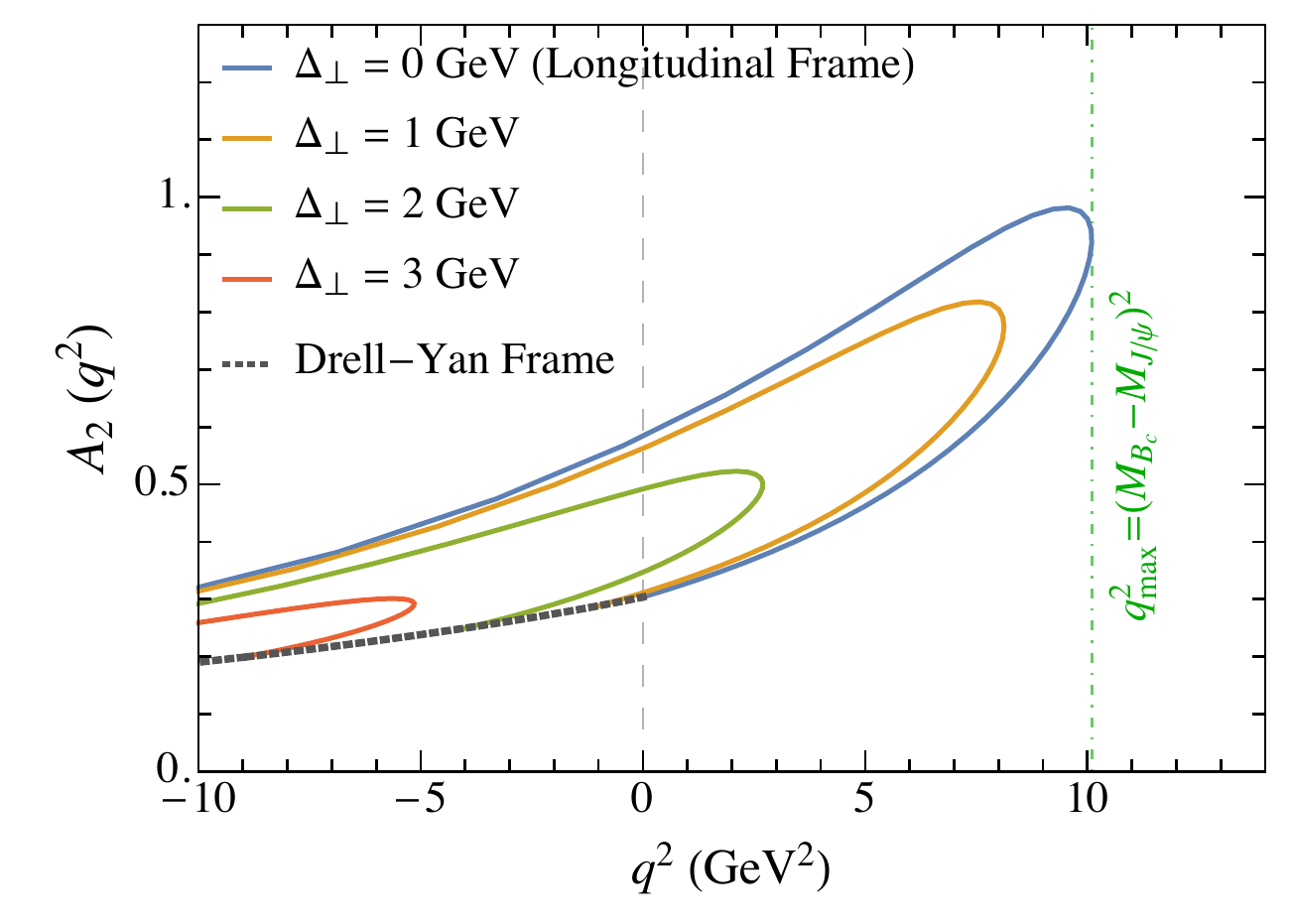}
		\end{subfigure}
		\hspace{1cm}
		\begin{subfigure}[t]{0.48\textwidth}
			\includegraphics[scale=0.65]{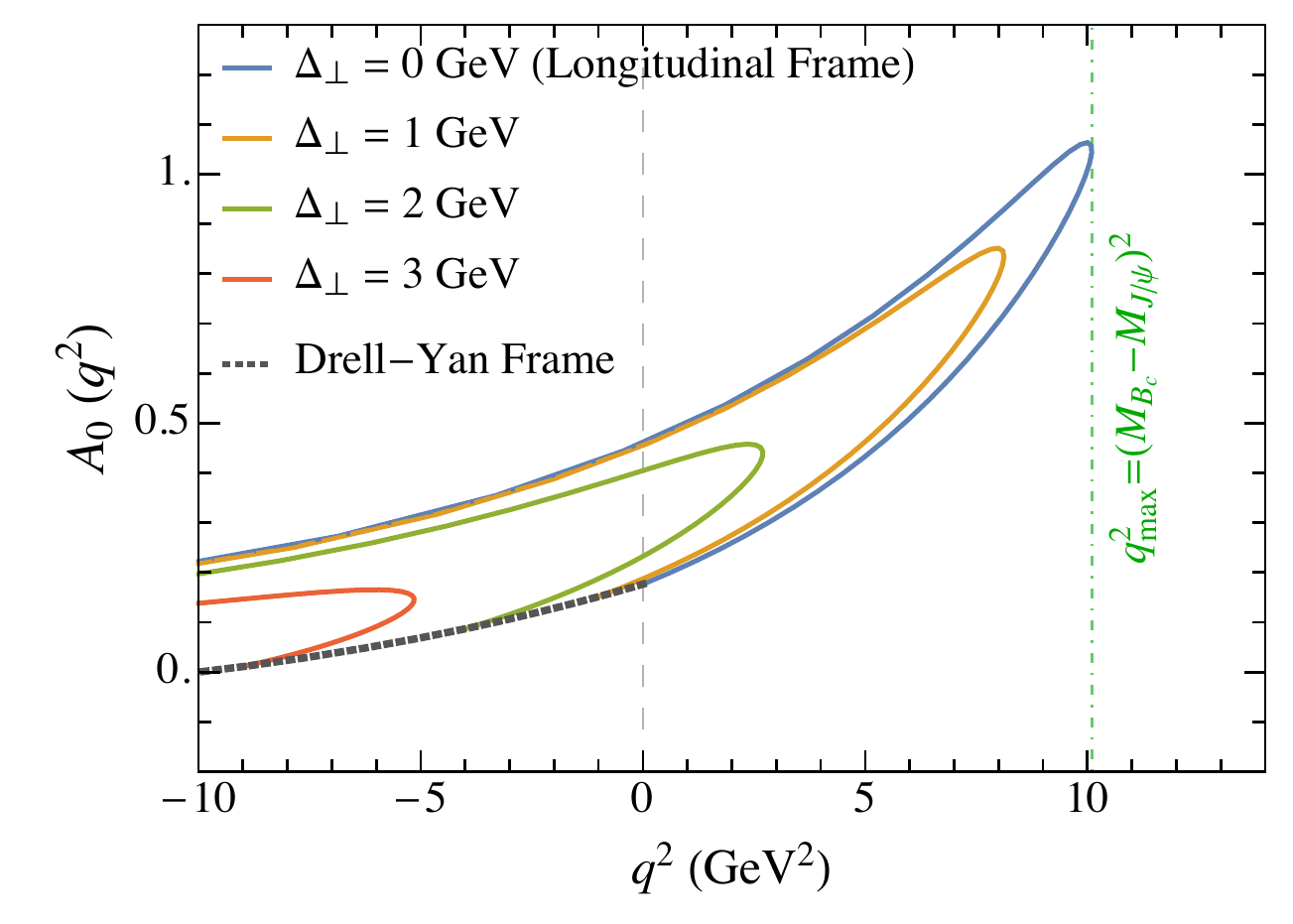}
		\end{subfigure}
		\caption{The frame dependence of form factors for $B_c \to J/\psi \ell \nu_\ell$ decay at the basis cutoff $N_\text{max} = L_\text{max} = 8$. Among the four form factors displayed here, $V(q^2)$ is only associated with a single hadronic matrix element and it shows a smaller frame dependence than the others.}
		\label{fig_frame_ffV}
	\end{figure*}

\begin{figure*}
		\hspace{-1cm}
		\begin{subfigure}[t]{0.48\textwidth}
			\includegraphics[scale=0.65]{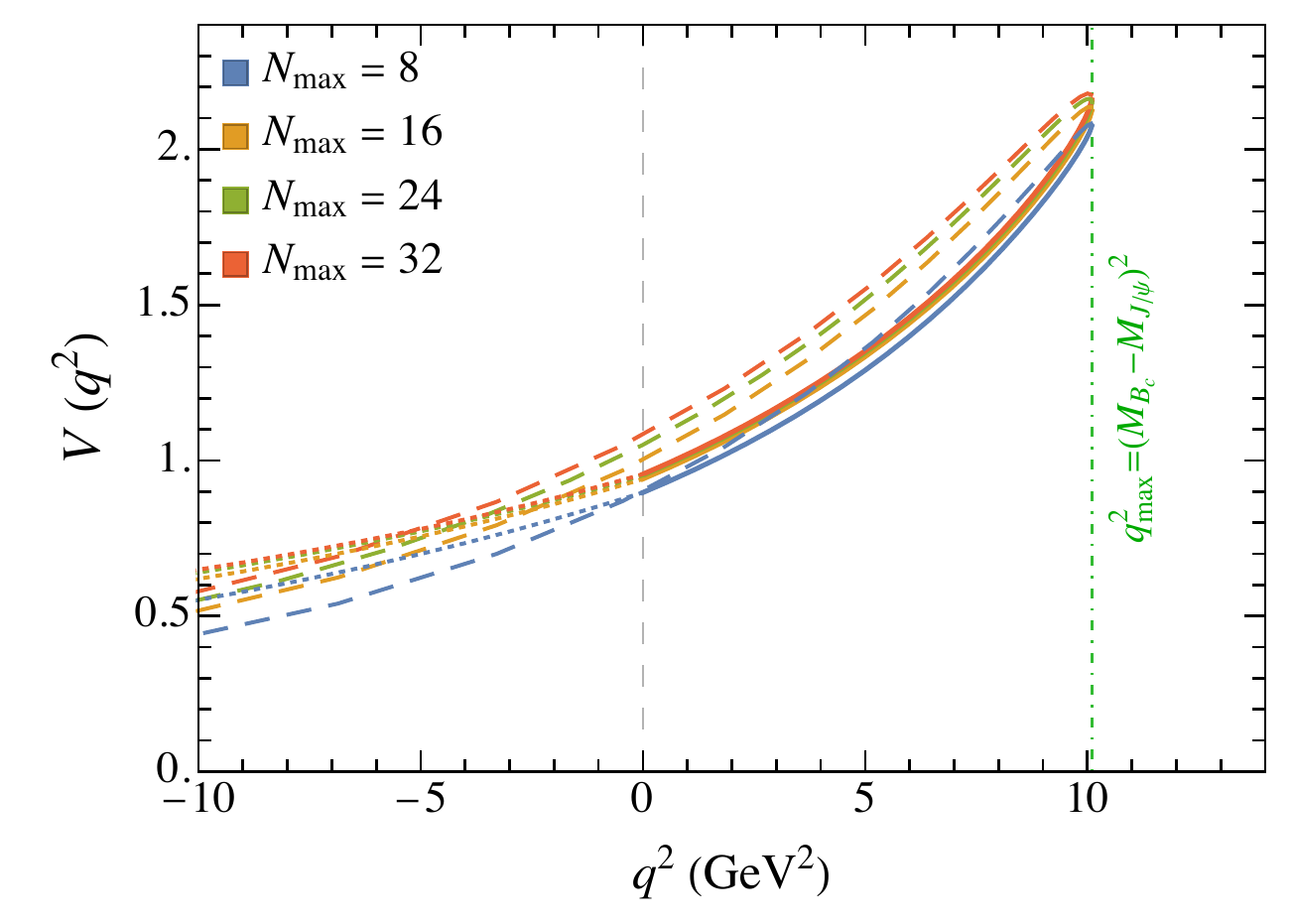}
		\end{subfigure}
		\hspace{1cm}
		\begin{subfigure}[t]{0.48\textwidth}
			\includegraphics[scale=0.65]{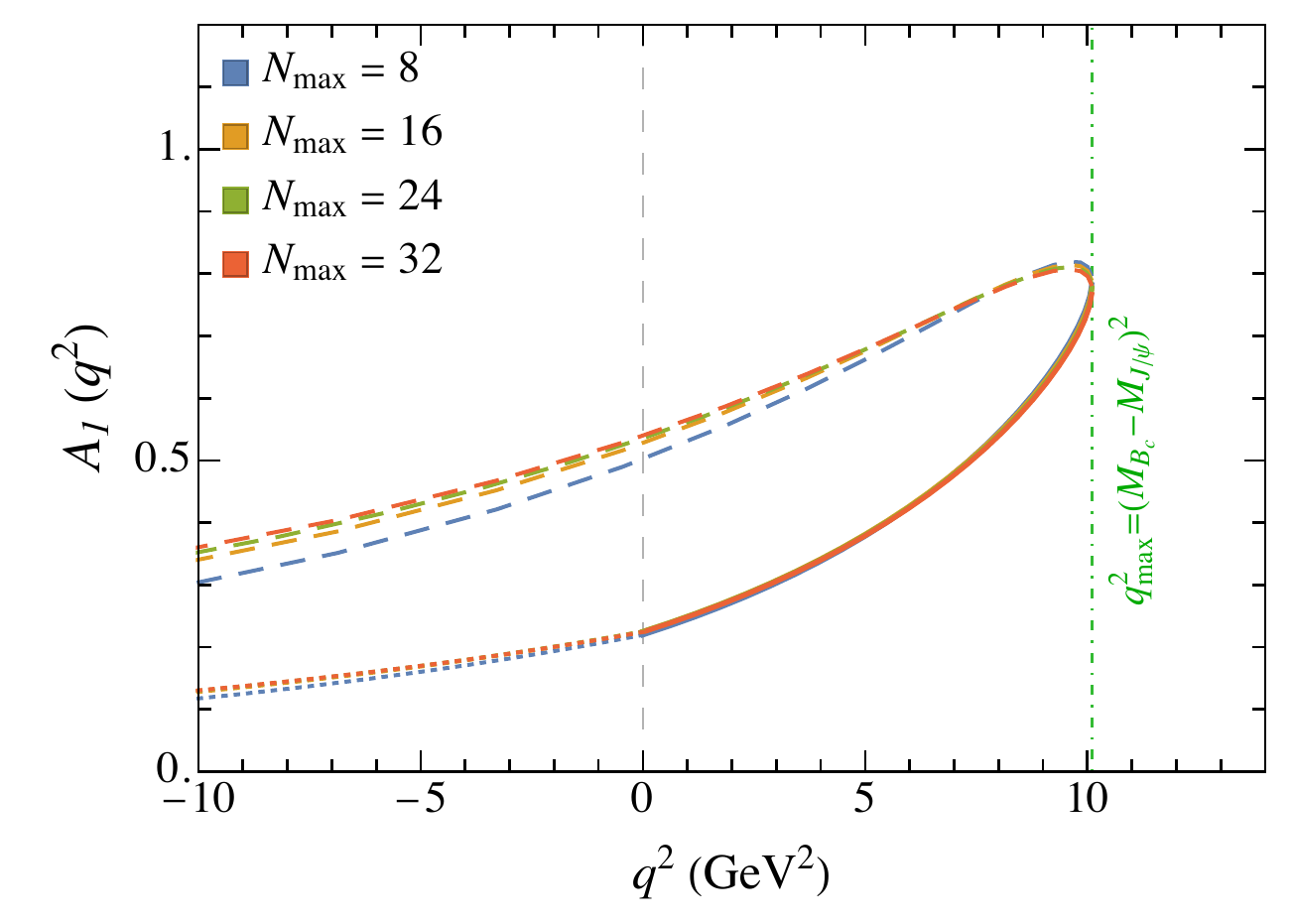}
		\end{subfigure}\\
		\hspace{-1cm}
		\begin{subfigure}[t]{0.48\textwidth}
			\includegraphics[scale=0.65]{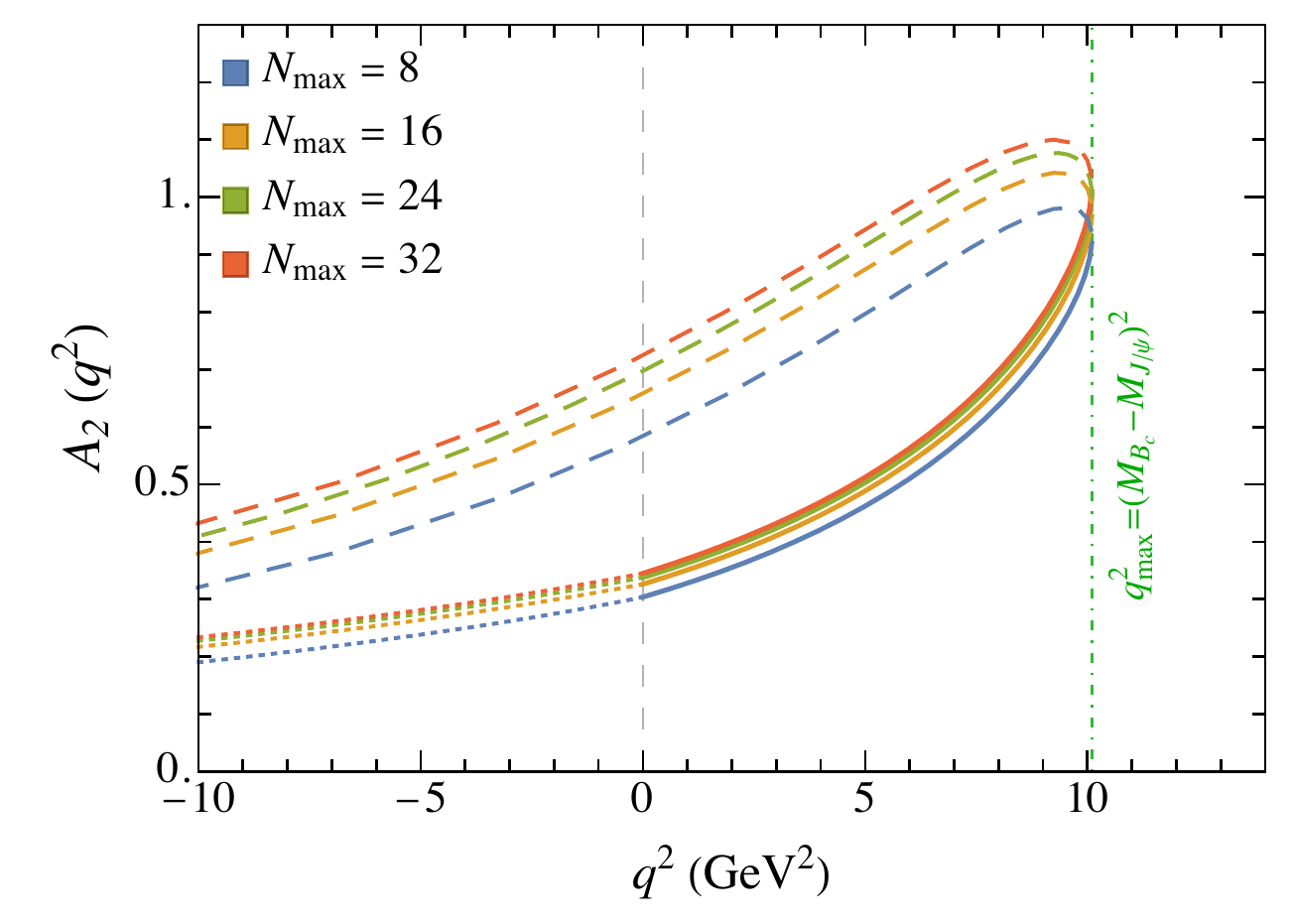}
		\end{subfigure}
		\hspace{1cm}
		\begin{subfigure}[t]{0.48\textwidth}
			\includegraphics[scale=0.65]{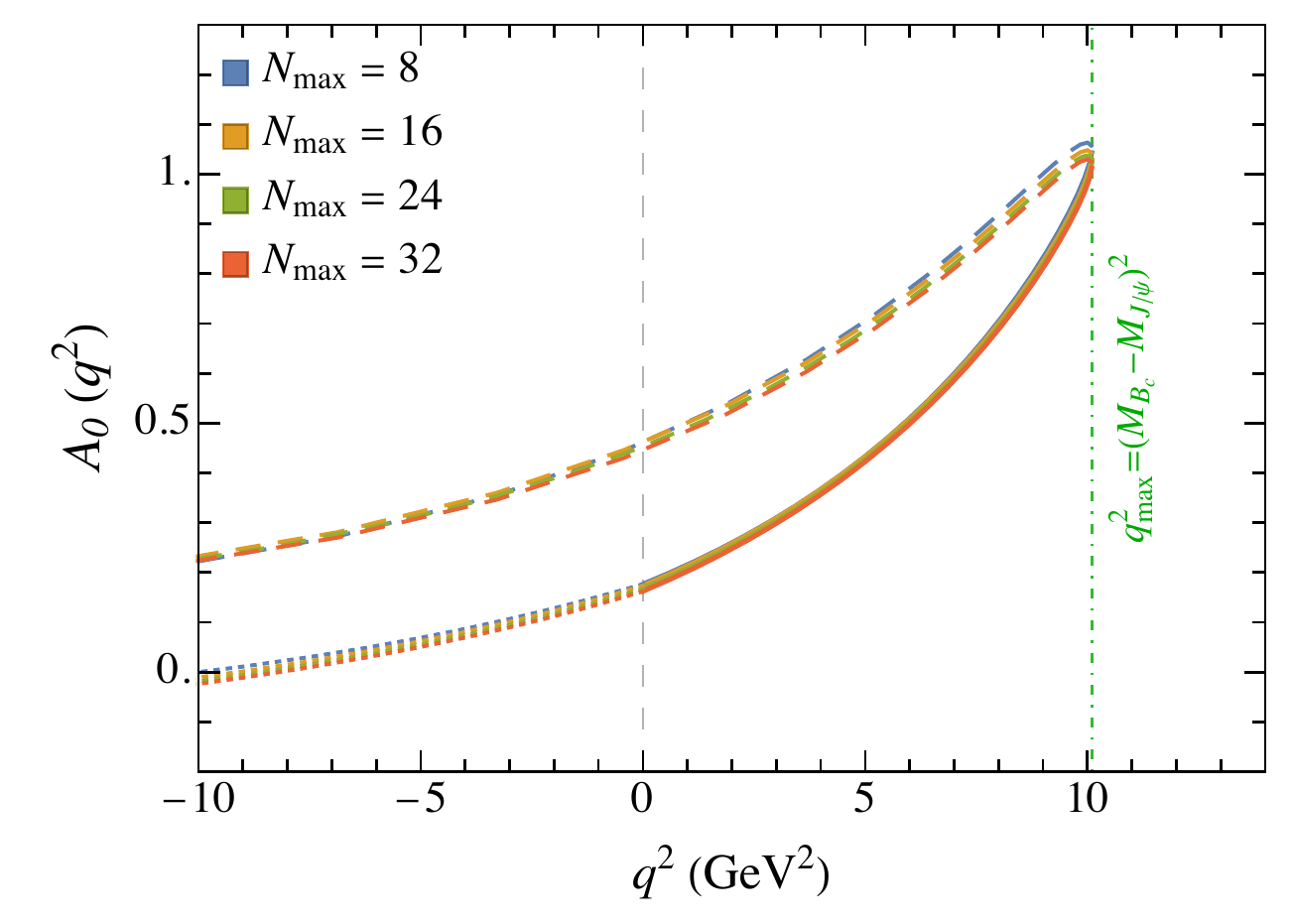}
		\end{subfigure}
		\caption{The dependence of the form factors on basis size as a function of $q^2$. Dotted lines are the results from Drell-Yan frame, solid curves are from the longitudinal-\textrm{I} frame while dashed curves are with the longitudinal-\textrm{II} frame.}
		\label{fig_basis_ffV}
	\end{figure*}

	Our results for the various form factors as functions of $q^2$ are shown in FIGs.~\ref{fig_frame_ffV} and~\ref{fig_basis_ffV}. The numerical results at $q^2=0$ and $q^2_\text{max}$ are listed in TABLE~\ref{tb2} in comparison with other approaches. 
	We notice that the frame dependence of the vector form factor $V(q^2)$, which is calculated using only one vector meson state and one current component, is significantly reduced compared to that of the axial vector form factors, which depend on two hadronic matrix elements involving different current components and different vector meson states.
	The frame dependence of $V(q^2)$ shows about $10\%$ deviation at $q^2=0$ which is consistent with Ref.~\cite{PhysRevD.100.036006};
	in fact, our results for $V(q^2)$ with the longitudinal-\textrm{II} is within about $15\%$ of the longitudinal-\textrm{I} and Drell-Yan frames over a wide range of $q^2$, both in the timelike and in the spacelike region.
	
\begin{figure}
		\centering
		\begin{subfigure}[t]{0.5\textwidth}
			\includegraphics[scale=0.48]{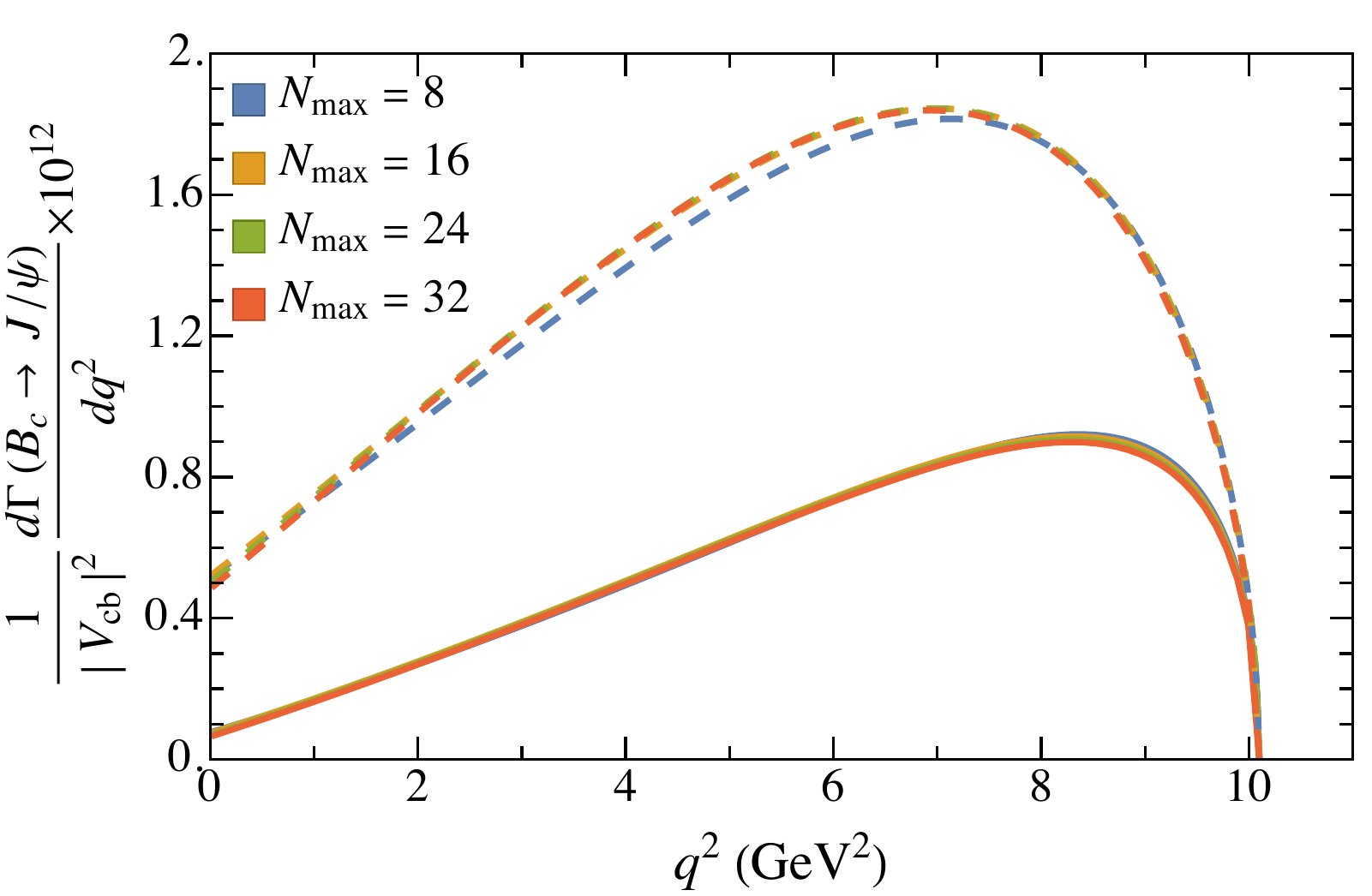}
		\end{subfigure}
		\caption{The differential decay width for the semileptonic decay of $B_c \to J/\psi e \bar{\nu}$. Results are presented with longitudinal-\textrm{I} (solid) and longitudinal-\textrm{II} (dashed) frames at different basis cutoffs.}
		\label{fig_widthV}
	\end{figure}

	The other three form factors ($A_1$, $A_2$, $A_0$) show behavior similar to $f_+(q^2)$ and $f_-(q^2)$ and exhibit a significantly stronger frame dependence than $V(q^2)$. When we further examine the basis dependence of form factors in FIG.~\ref{fig_basis_ffV}, we find only very modest sensitivity to the basis truncation, much less than the form factors $f_\pm(q^2)$ in the previous subsection. A similar insensitivity to basis space cutoff is found for the differential decay width of $B_c \to J/\psi$ as shown in FIG.~\ref{fig_widthV}. Unlike the decay width for $B_c \to \eta_c$ where the longitudinal-\textrm{I} results vary by as much as $100\%$, this decay width with a vector meson final state does not change substantially with basis truncation, leaving the frame dependence as the dominant source of variations. Similar to the case with pseudoscalar final states, we keep our preference in results based on the longitudinal-\textrm{I} frame for these from factors with a vector meson final state.
	
	\begin{table*}
		\begin{tabular}{c|cccccccccc}
			\toprule
			&  \hspace{0.1cm}  	\multirow{2}{*}{ $V(0)$ }  \hspace{0.15cm} &  \hspace{0.15cm}  	\multirow{2}{*}{ $V(q^2_\text{max})$} \hspace{0.15cm}  &  \hspace{0.15cm} 	\multirow{2}{*}{ $A_1(0)$}  \hspace{0.15cm} &   \hspace{0.15cm} 	\multirow{2}{*}{ $A_1(q^2_\text{max})$}   \hspace{0.15cm} 
			&  \hspace{0.15cm}  	\multirow{2}{*}{ $A_2(0)$ }  \hspace{0.15cm} &  \hspace{0.15cm}  	\multirow{2}{*}{ $A_2(q^2_\text{max})$} \hspace{0.15cm}  &  \hspace{0.15cm} 	\multirow{2}{*}{ $A_0(0)$}  \hspace{0.15cm} &   \hspace{0.15cm} 	\multirow{2}{*}{ $A_0(q^2_\text{max})$}   \hspace{0.1cm}\\
			\\
			\midrule
			BLFQ - 1& $0.956(8)	$	& \multirow{2}{*}{$2.166(29)$} & $0.224(2)$ & \multirow{2}{*}{$0.773(7)$}   &$0.345(15)$&   \multirow{2}{*}{$1.020(36)$} & $0.162(10)$ &  \multirow{2}{*}{$1.017(14)$}  \\
			BLFQ - 2& $1.082(64)$	& & $0.540(8)$ & &$0.724(54)$ && $0.445(16)$ & \\
			pQCD~\cite{Wang_2013} & 0.42 & 0.94 & 0.46&0.79 &0.64 & 1.86 & 0.52 & 0.99  \\
			CCQM~\cite{ISSADYKOV2018178} & 0.78  & 1.32 &0.56&0.79 &0.55& 0.89& 0.56&  0.82 \\
			RQM~\cite{PhysRevD.68.094020} & 0.49& 1.34 & 0.50 &0.88 &0.73 & 1.33& 0.40& 1.06 \\
			Lattice QCD~\cite{Colquhoun:2016osw} & 0.70& &0.48&\\
			\bottomrule
		\end{tabular}
		\caption{Form factors by this work (rows labeled BLFQ - 1 and BLFQ - 2) and other methods at selected values of $q^2$. The BLFQ results listed here are calculated with $N_\text{max}=L_\text{max}=32$, and the uncertainties are induced by the same source as in TABLE~\ref{tb1}.}
		\label{tb2}
	\end{table*}

	For further investigation, we integrate the differential decay width over $q^2$ in the physical region $m_l^2 \le q^2 \le q^2_\text{max}$ to obtain the corresponding total decay width and, from that, the branching ratio. 
	TABLE~\ref{tb3} is a comparison of our results with other calculations from Refs.~\cite{Wang_2013,ISSADYKOV2018178,PhysRevD.87.014009,Dhir_2010,yao2021semileptonic,Rui2016,PhysRevD.68.094020,PhysRevD.61.034012,chang2015some} for the semileptonic decays into $\eta_c$ and $J/\psi$, as well as the decays into the $2S$ radial excited states $\eta'_c$ and $\psi'$. The following supporting physical parameters including $\abs{V_{cb}}$ are adopted from PDG~\cite{10.1093/ptep/ptaa104}:
	\begin{equation}
		\begin{gathered}
			m_\mu = 105.6583745 \text{ MeV;} \quad m_\tau =1776.86 \text{ MeV;} \\
			\abs{V_{cb}} = 41.0 \times 10^{-3}; \quad \tau_{B_c} = 0.510 \times 10^{-12} \text{s}.  \\
		\end{gathered}
	\end{equation}
	The decay modes with $e$ and $\mu$ in the final products have negligible difference in branching ratios, while the decay modes with a $\tau$ in the final states are suppressed due to limits in the phase space.
	Despite the difference between BLFQ - 1 and BLFQ - 2 results, we note that there exist other methods that appear compatible with either of our BLFQ results. Specifically, we found that the BLFQ - 1 results for the decays into pseudoscalar states have better agreement with other methods than BLFQ - 2, which is in accord with our preference for the longitudinal-\textrm{I} frame. 
	However, for the decays into vector states, neither frame seems to be superior when compared to other calculations.  Nevertheless, also for the branching ratios into vector states, we expect our BLFQ - 1 results to be more reliable than our BLFQ - 2 results.

	\begin{table*}
		\centering 
		\begin{tabular}{c|cccc}
			\toprule
			\multirow{2}{*}{ $\mathcal{BR}$}  & 	\multirow{2}{*}{ $B_c \rightarrow \eta_c l \bar{\nu}_l\ ( l = e, \mu)$ } & 	\multirow{2}{*}{ $B_c \rightarrow \eta_c \tau \bar{\nu}_\tau $ }& 
			\multirow{2}{*}{ $B_c \rightarrow J/\psi l \bar{\nu}_l\ ( l = e, \mu)$ } & 	\multirow{2}{*}{  $B_c \rightarrow J/\psi \tau \bar{\nu}_\tau$ } \\
			\\
			\midrule
			BLFQ - 1 & $0.722\pm 0.034$	&  $0.188 \pm 0.006$	 & $0.727\pm0.014$  &$0.228\pm0.004$	\\
			BLFQ - 2 &  $1.88\pm0.39$	&  $0.404\pm0.06$& $1.75\pm0.02$  & $0.452\pm0.006$\\
			pQCD\cite{Wang_2013} & $0.441$ & $0.137$ & $1.003$ & $0.292$ \\
			CCQM\cite{ISSADYKOV2018178} &$0.96\pm 0.19$&$0.24 \pm 0.05$&$1.67\pm 0.33$&$0.40\pm0.08$\\
			NRQCD\cite{PhysRevD.87.014009} & $2.1^{+0.5+0.4+0.2}_{-0.3-0.1-0.2}$ & $0.64^{+0.07+0.14+0.10}_{-0.08-0.06-0.05}$ & $6.7^{+0.07+0.14+0.10}_{-0.08-0.06-0.05}$ & $ 0.52^{+0.16+0.08+0.08}_{-0.09-0.03-0.05}$ \\
			RQM\cite{Dhir_2010} 	&	$0.606^{+0.035}_{-0.025}$	&	$0.195^{+0.011}_{-0.007}$	&	$1.10^{+0.06}_{-0.09}$	&	$0.264^{+0.022}_{-0.016}$\\
			CSM \cite{yao2021semileptonic} 	&	$0.810\pm0.045\pm0.055$	&	$0.254\pm0.010\pm0.017$	&	$1.72\pm0.019\pm0.012$	&	$0.417\pm0.066\pm0.028$\\
			\bottomrule
			\end{tabular}
			\vspace{1cm}
			\begin{tabular}{c|cccc}
			\toprule
			\multirow{2}{*}{ $\mathcal{BR}$}  & 	\multirow{2}{*}{ $B_c \rightarrow \eta'_c l \bar{\nu}_l\ ( l = e, \mu)$ } & 	\multirow{2}{*}{ $B_c \rightarrow \eta'_c \tau \bar{\nu}_\tau $ }& 
			\multirow{2}{*}{ $B_c \rightarrow \psi' l \bar{\nu}_l\ ( l = e, \mu)$ } & 	\multirow{2}{*}{  $B_c \rightarrow \psi' \tau \bar{\nu}_\tau$ } \\
			\\
			\midrule
			BLFQ - 1 & $0.0371\pm0.0001$	&  $1.36\pm 0.02 \times 10^{-3}$	 & $ 0.0710\pm0.0036$  &$ 4.86\pm 0.421 \times10^{-3}$	\\
			BLFQ - 2 &  $0.315\pm0.050$	&  $6.91\pm 0.83 \times 10^{-3} $& $0.0442 \pm 0.0081$  & $ 3.12 \pm 0.592\times10^{-3}$\\
			pQCD\cite{Rui2016} &	$0.77^{+0.20+0.58+0.20}_{-0.14-0.55-0.05}$	&	$5.3^{+1.4+4.1+1.4}_{-1.0-3.8-0.3} \times 10^{-2}$		&	$1.2^{+0.6+0.1+0.3}_{-0.3-0.1-0.1}$	&	$8.4^{+3.6+0.4+1.5}_{-1.3-0.4-0.1} \times 10^{-2} $	\\
			RQM \cite{PhysRevD.68.094020}& $0.032$ & & $0.031$ &  \\
			RQM \cite{PhysRevD.61.034012}& $0.02$ & & $0.12$ &  \\
			BSE\cite{chang2015some}&$0.0665^{+0.0052}_{-0.0060}$& & $0.103^{+0.013}_{-0.018}$& \\
			\bottomrule
		\end{tabular}
		\caption{Branching ratios (in $\%$) of semileptonic $B_c$ decays into $\eta_c$ and $J/\psi$ and the radial excited states $\eta'_c$ and $\psi'$. The branching ratio ($\mathcal{BR}$) central values for BLFQ are calculated with $N_\text{max}=L_\text{max}=32$, while the uncertainties are given by $\varepsilon_{\mathcal{BR}} = 2 \abs{\mathcal{BR}(N_\text{max}=32) - \mathcal{BR}(N_\text{max}=24)}$ to show the sensitivity to basis cutoffs. This work is compared with other frameworks including the non-relativistic QCD (NRQCD) approach, the continuum Schwinger function (CSM) method, and the Bethe-Salpeter equation (BSE) method. Note that in some references, slightly different values of $V_{cb}$ and $\tau_{B_c}$ are adopted for the calculation.}
		\label{tb3}
	\end{table*}
	
	\section{Summary}			
	\label{sec_summary}	
	We investigated the semileptonic decay form factors, differential decay width, and the branching ratios of the $B_c$ meson to the pseudoscalar state $\eta_c$ and vector state $J/\psi$ with the BLFQ approach. 
	In order to access the electroweak form factors in the timelike region using light-front kinematics, we introduced two boost invariants $z$ and $\vec{\Delta}_\perp$ that specify the choice of reference frames. The frame dependence of form factors arises due to our Fock space truncation. We expect our results in the Drell-Yan and longitudinal-\textrm{I} frames to be more reliable than those in the longitudinal-\textrm{II} frame because the contribution from the particle-number-changing diagrams, which are not incorporated in our truncated Fock space, is suppressed in the frames with small longitudinal momentum transfer. 
	
	We employ the LFWFs obtained in our previous work~\cite{PhysRevD.96.016022,PhysRevD.98.114038} which are available for certain basis sizes. Within the Hilbert space truncated by the basis cutoffs, we observed that the influence on the form factors due to the basis truncation is smaller than the frame dependence. This conclusion is supported best in the Drell-Yan and longitudinal-\textrm{I} frames, which coincide with the ones minimizing the error due to omitting the particle-number-changing diagram. 
	In general, when comparing our results of form factors and branching ratios to those in the literature (see TABLES~\ref{tb2} and~\ref{tb3}), we find reasonable agreement but some differences are noticeable.
	Further calculations of the weak transition form factors can be repeated for other weak decay processes involving $B^{(*)}$, $B_s^{(*)}$, $D^{(*)}$, and $D_s^{(*)}$~\cite{PhysRevLett.120.171802,PhysRevD.101.072004} with the corresponding LFWFs~\cite{Tang:2019gvn}. Also note that for the decay into $J/\psi$, we select a certain combination of current components and final state magnetic projection as specified in Eq.~\eqref{eq_ffvec}. It is worthwhile to test other combinations in a future work and investigate the sensitivity to the choices of current and magnetic projection.
	
	 Our preference of the Drell-Yan and the longitudinal-\textrm{I} frames is linked with the Fock space of the LFWFs being limited to the valence quarks. With BLFQ for mesons incorporating higher Fock sectors in the future, we expect a reduction of the existing artifact of frame dependence. We anticipate future experimental facilities such as LHCb to provide more data to test our results and motivate further theoretical improvements.

	\section{Acknowledgments}			
	We wish to thank Yang Li, Meijian Li, Wenyang Qian, and Anji Yu for valuable discussions. This work was supported in part by the U.S. Department of Energy under Grants No. DE-FG02-87ER40371 and No. DE-SC0018223 (SciDAC-4/NUCLEI). Computational resources were provided by the National Energy Research Supercomputer Center, which is supported by the Office of Science of the U.S. Department of Energy under Contract No. DE-AC02-05CH11231. S.~J. is also supported by U.S. Department of Energy, Office of Science, Office of Nuclear Physics, contract no. DE-AC02-06CH11357.

\appendix
\section{Basis representation}
\label{appxA}
The effective Hamiltonian we employed for the mesons include the HO transverse confining potential, a longitudinal confinement, and an effective one-gluon exchange~\cite{LI2016118,PhysRevD.96.016022}
\begin{equation}
	\begin{aligned}
		& H_\text{eff}  = \frac{\vec{k}^2_\perp+m_q^2}{x} + \frac{\vec{k}^2_\perp+m_{\bar{q}}^2}{1-x} + \kappa^4\vec{\zeta}^2_\perp  \\
		& \quad - \frac{\kappa^4}{(m_q+m_{\bar{q}})^2} \partial_x (x(1-x)\partial_x)\\
		& \quad -\frac{C_F 4 \pi \alpha_s(Q^2)}{Q^2} \bar{u}_{s'}(k')\gamma_\mu  u_s(k) \bar{v}_{\bar{s}}(\bar{k}) \gamma^\mu v_{\bar{s}'}(\bar{k}').
	\end{aligned}
\end{equation}
Diagonalizing this Hamiltonian in our chosen basis space provides the eigenvalues as squares of the bound state eigenmasses, and the eigenvectors $\psi_h$ which specify the LFWFs as
\begin{equation}
	\begin{aligned}
		\psi_{s\bar{s}/h}^{(m_j)} \left(x, \vec{k}_\perp\right) & = \sum_{nml}\, \psi_h(n,m,l,s,\bar{s})\chi_l(x)  \\ 
		& \quad \times \phi_{nm}(\vec{k}_\perp/\sqrt{x(1-x)}).
	\end{aligned}
\end{equation}
Here $\phi_{nm}(\vec{q}_\bot)$ is the 2D HO function we adopt as the transverse basis function:
\begin{equation}
	\begin{aligned}
		\phi_{nm} (\vec{q}_\perp) & = \frac{1}{\kappa}\sqrt{\frac{4\pi n!}{(n+\abs{m})!}} \Big(\frac{q_\perp}{\kappa}\Big) ^{\abs{m}} e^{-\frac{1}{2} q^2_\perp / \kappa^2}  \\
		& \quad \times 
		L_n^{\abs{m}} (q^2_\perp/\kappa^2) e^{\imag m \theta_q},
	\end{aligned}
\end{equation}
where $n$ and $m$ are the principal and orbital quantum numbers, respectively; $\kappa$ is the scale parameter, $\theta_q=\arg \vec{q}_\perp$, and $L_n^{\abs{m}}$ is the associated Laguerre polynomial.
The longitudinal basis function $\chi_l(x)$ is related to the Jacobi polynomial with quantum number $l$ by
\begin{equation}
	\begin{aligned}
		\chi_l(x) & = \sqrt{4\pi(2l+\alpha+\beta+1)} \\
		& \quad \times \sqrt{\frac{\Gamma(l+1)\Gamma(l+\alpha+\beta+1)}{\Gamma(l+\alpha+1)\Gamma(l+\beta+1)}}  \\
		& \quad \times  x^{\frac{\beta}{2}}(1-x)^{\frac{\alpha}{2}} P_l^{(\alpha,\beta)}(2x-1),\\
	\end{aligned}
\end{equation}
where $\alpha$ and $\beta$ are two parameters associated with quark masses, i.e. $\alpha={2 m_{\bar{q}}\left(m_{q}+m_{\bar{q}}\right) / \kappa^2}$, ${\beta=2 m_{q}\left(m_{q}+m_{\bar{q}}\right) / \kappa^2}$.
The basis truncation is specified by the dimensionless parameters $N_\text{max}$ and $L_\text{max}$, such that 
\begin{equation}
	2n+\abs{m}+1 \le N_\text{max},\qquad 0 \le l \le L_\text{max}.
\end{equation}

One can then write down the hadronic matrix element in the basis representation. Here we take $\mathcal{M}^\mu = \mel{P_2}{\bar{c}\gamma^\mu b}{P_1}$ for $B_c \to \eta_c$ as an example:
\begin{widetext}
	\begin{equation}
		\begin{aligned}
			\mathcal{M}^\mu =& \sum_{nn'mm'll's \bar{s}} \psi_{\eta_c}(n',m',l',s', \bar{s}) \psi_{B_c}(n,m,l,s, \bar{s}) 
			\int_z^1 \frac{\dd x}{2x(1-x)} \int\frac{\dd ^2 \vec{k}_\perp}{(2\pi)^3} \sum_{s'} \frac{1-z}{x-z} \\
			&\times \bar{u}_{s'} \left( (x-z)P_1^+, \vec{k}_\perp + (x-z)\vec{P}_{1\perp} -\vec{\Delta}_\perp \right) \gamma^\mu u_s \left(x P_1^+,\vec{k}_\perp+x \vec{P}_{1\perp} \right)\\
			&\times \phi^*_{n'm'}\left(\frac{\vec{k}_\perp-(\frac{1-x}{1-z})\vec{\Delta}_\perp}{\sqrt{(x-z)(1-x)}/(1-z)}\right) \chi_{l'}\left( \frac{x-z}{1-z}\right) \phi_{nm}\left(\frac{\vec{k}_\perp}{\sqrt{x(1-x)}}\right) \chi_l(x).
		\end{aligned}
	\end{equation}
\end{widetext}
Note that the basis states for the initial and final mesons have different values for their constants owing to the dependence on quark masses and interaction parameters in their respective Hamiltonians.

\section{Spinor vertices}
\label{appxB}
We present the spinor vertex, where $u_s(p)$ and $\bar{u}_{s'}(p')$ are the solutions of the Dirac equation for the incoming quark and outgoing antiquark with masses $m_1$ and $m_2$, respectively.
\paragraph{Vector}
\begin{subequations}
	\begin{align}
		& \bar{u}_{s'}(p') \gamma^+ u_s(p) =2\sqrt{p^+p'^+} \delta_{ss'}\quad , 
	\end{align}
	\begin{align}
		& \bar{u}_{s'}(p') \gamma^- u_s(p) =\frac{2}{\sqrt{p^+p'^+}}  \nonumber\\
		& \quad \times \left\{
		\begin{aligned}
			m_1 m_2 + p^R p'^L \qquad &s,s' = +,+ \\
			m_1 m_2 + p^L p'^R \qquad &s,s' = -,- \\
			m_2p^R-m_1p'^R  \qquad &s,s' = +,- \\
			m_1p'^L -m_2p^L \qquad &s,s' = -,+ 
		\end{aligned}\quad,
		\right.
	\end{align}
	\begin{align}
		& \bar{u}_{s'}(p') \gamma^L u_s(p) =2\sqrt{p^+p'^+} \nonumber \\
		& \quad \times \left\{
		\begin{aligned}
			\frac{p'^L}{p'^+} \qquad\qquad\quad &s,s' = +,+ \\
			\frac{p^L}{p^+}  \qquad\qquad\quad &s,s' = -,- \\
			\frac{m_2}{p'^+} - \frac{m_1}{p^+}   \qquad &s,s' = +,- \\
			0 \qquad\qquad\qquad &s,s' = -,+ 
		\end{aligned}\quad,
		\right.
	\end{align}
	\begin{align}
		& \bar{u}_{s'}(p') \gamma^R u_s(p) =2\sqrt{p^+p'^+}  \nonumber\\
		& \quad \times \left\{
		\begin{aligned}
			\frac{p^R}{p^+} \qquad\qquad\quad &s,s' = +,+ \\
			\frac{p'^R}{p'^+}  \qquad\qquad\quad &s,s' = -,- \\
			0  \qquad\qquad\qquad &s,s' = +,- \\
			\frac{m_1}{p^+} - \frac{m_2}{p'^+} \qquad &s,s' = -,+ 
		\end{aligned}\quad.
		\right. 
	\end{align}
\end{subequations}

\paragraph{Axial vector}
\begin{subequations}
	\begin{align}
		\bar{u}_{s'}(p') \gamma^+ \gamma_5 u_s(p) &=2\sqrt{p^+p'^+}   \delta_{ss'} \text{sign}(s)\quad,
	\end{align}
	\begin{align}
		& \bar{u}_{s'}(p') \gamma^- \gamma_5u_s(p) =\frac{2}{\sqrt{p^+p'^+}}  \nonumber\\
		& \quad \times \left\{
		\begin{aligned}
			- m_1 m_2 + p^R p'^L \qquad &s,s' = +,+ \\
			m_1 m_2 - p^L p'^R \qquad &s,s' = -,- \\
			m_2p^R+m_1p'^R  \qquad &s,s' = +,- \\
			m_1p'^L +m_2p^L \qquad &s,s' = -,+ 
		\end{aligned}\quad,
		\right.
	\end{align}
	\begin{align}
		& \bar{u}_{s'}(p') \gamma^L \gamma_5 u_s(p) = 2\sqrt{p^+p'^+}  \nonumber\\
		& \quad \times \left\{
		\begin{aligned}
			\frac{p'^L}{p'^+}\qquad\qquad\quad &s,s' = +,+ \\
			-\frac{p^L}{p^+}  \qquad\qquad\  &s,s' = -,- \\
			\frac{m_1}{p^+} +\frac{m_2}{p'^+} \qquad &s,s' = +,- \\
			0 \qquad\qquad\qquad &s,s' = -,+ 
		\end{aligned}\quad,
		\right.
	\end{align}
	\begin{align}
		& \bar{u}_{s'}(p') \gamma^R \gamma_5 u_s(p) =2\sqrt{p^+p'^+} \nonumber \\
		& \quad \times \left\{
		\begin{aligned}
			\frac{p^R}{p^+} \qquad\qquad\quad &s,s' = +,+ \\
			-\frac{p'^R}{p'^+}  \qquad\qquad\  &s,s' = -,- \\
			0  \qquad\qquad\qquad &s,s' = +,- \\
			\frac{m_1}{p^+} + \frac{m_2}{p'^+} \qquad &s,s' = -,+ 
		\end{aligned}\quad.
		\right.
	\end{align}
\end{subequations}

\bibliography{weakdcy}
\end{document}